\documentclass[Afour,sageh,times,doublespace]{sagej}

\usepackage{moreverb,url}

\usepackage{graphicx}
\usepackage{amssymb}
\usepackage{amsmath}
\usepackage[acronym]{glossaries}
\usepackage{float}
\usepackage{pgfplots}
\pgfplotsset{compat=1.16}
\usepackage{breqn}
\usepackage[colorlinks,bookmarksopen,bookmarksnumbered,citecolor=red,urlcolor=red, hidelinks]{hyperref}
\usepackage{interval}
\usepackage{tikz}
\usetikzlibrary{patterns,shapes.arrows,calc,angles,quotes,babel,arrows,arrows.meta,shapes.geometric,svg.path}
\usepackage[capitalise]{cleveref}
\usepackage{xurl}
\usepackage{uri}
\usepackage{subfig}
\usepackage{scalerel}

\definecolor{orcidlogocol}{HTML}{A6CE39}
\tikzset{
  orcidlogo/.pic={
    \fill[orcidlogocol] svg{M256,128c0,70.7-57.3,128-128,128C57.3,256,0,198.7,0,128C0,57.3,57.3,0,128,0C198.7,0,256,57.3,256,128z};
    \fill[white] svg{M86.3,186.2H70.9V79.1h15.4v48.4V186.2z}
                 svg{M108.9,79.1h41.6c39.6,0,57,28.3,57,53.6c0,27.5-21.5,53.6-56.8,53.6h-41.8V79.1z M124.3,172.4h24.5c34.9,0,42.9-26.5,42.9-39.7c0-21.5-13.7-39.7-43.7-39.7h-23.7V172.4z}
                 svg{M88.7,56.8c0,5.5-4.5,10.1-10.1,10.1c-5.6,0-10.1-4.6-10.1-10.1c0-5.6,4.5-10.1,10.1-10.1C84.2,46.7,88.7,51.3,88.7,56.8z};
  }
}

\newcommand\orcidicon[1]{\href{https://orcid.org/#1}{\mbox{\scalerel*{
\begin{tikzpicture}[yscale=-1,transform shape]
\pic{orcidlogo};
\end{tikzpicture}
}{|}}}}

\definecolor{matplotlibBlue}{HTML}{1f77b4} 
\definecolor{matplotlibOrange}{HTML}{ff7f0e} 
\definecolor{matplotlibGreen}{HTML}{2ca02c} 
\definecolor{matplotlibRed}{HTML}{d62728} 
\definecolor{matplotlibPurple}{HTML}{9467bd} 
\definecolor{matplotlibBrown}{HTML}{8c564b} 
\definecolor{matplotlibPink}{HTML}{e377c2} 
\definecolor{matplotlibGrey}{HTML}{7f7f7f} 
\definecolor{matplotlibYellow}{HTML}{bcbd22} 

\tikzstyle{kernel} = [rectangle, 
minimum width=3cm, 
minimum height=1cm, 
text centered, 
text width=3cm, 
draw=black]

\tikzstyle{syncSweep} = [rectangle, 
minimum width=3cm, 
minimum height=1cm, 
text centered, 
text width=3cm, 
draw=black, 
fill=matplotlibBlue!50]

\tikzstyle{syncKernel} = [rectangle, 
minimum width=3cm, 
minimum height=1cm, 
text centered, 
text width=3cm, 
draw=black, 
fill=matplotlibOrange!50]

\tikzstyle{hybridKernel} = [rectangle, 
minimum width=3cm, 
minimum height=1cm, 
text centered, 
text width=3cm, 
draw=black, 
fill=matplotlibGreen!50]

\tikzstyle{decision} = [diamond, 
minimum width=3cm, 
minimum height=1cm, 
text centered, 
draw=black]
\tikzstyle{arrow} = [thick,->,>=stealth]

\newcommand{\walberla}{\textsc{waLBerla}}
\newcommand{\change}[1]{{\color{black} {#1}}}

\newcommand\BibTeX{{\rmfamily B\kern-.05em \textsc{i\kern-.025em b}\kern-.08em
T\kern-.1667em\lower.7ex\hbox{E}\kern-.125emX}}

\def\volumeyear{2024}

\begin{document}

\def\journalname{The International Journal of High Performance Computing Applications}

\runninghead{Kemmler et al.}

\title{Efficiency and scalability of fully-resolved fluid-particle simulations on heterogeneous CPU-GPU architectures}

\author{Samuel Kemmler\affilnum{1,2}\orcidicon{0000-0002-9631-7349}, Christoph Rettinger\affilnum{1}\orcidicon{0000-0002-0605-3731}, Ulrich Rüde\affilnum{1,3}\orcidicon{0000-0001-8796-8599},\newline Pablo Cuéllar\affilnum{2}\orcidicon{0000-0003-2446-8065} and Harald Köstler\affilnum{1}\orcidicon{0000-0002-6992-2690}}

\affiliation{\affilnum{1}Chair for System Simulation, Friedrich-Alexander-Universität Erlangen-Nürnberg, Erlangen, Germany\\\affilnum{2}Division 7.2 for Buildings and Structures, Federal Institute for Materials Research and Testing (BAM), Berlin, Germany\\\affilnum{3}CERFACS, Toulouse, France}

\corrauth{Samuel Kemmler, Chair for System Simulation, Friedrich-Alexander-Universität Erlangen-Nürnberg, Cauerstraße 11, 91058 Erlangen, Germany}

\email{samuel.kemmler@fau.de}

\newacronym{lbm}{LBM}{Lattice Boltzmann Method}
\newacronym{dem}{DEM}{Discrete Element Method}
\newacronym{cpu}{CPU}{Central Processing Unit}
\newacronym{gpu}{GPU}{Graphics Processing Unit}
\newacronym{psm}{PSM}{Partially Saturated Cells Method}
\newacronym{mem}{MEM}{Momentum Exchange Method}
\newacronym{pdfs}{PDFs}{Particle Distribution Functions}
\newacronym{cfd}{CFD}{Computational Fluid Dynamics}
\newacronym{simd}{SIMD}{Single Instruction, Multiple Data}
\newacronym{bc}{BC}{Boundary Condition}
\newacronym{srt}{SRT}{Single Relaxation Time}
\newacronym{pd}{PD}{Particle Dynamics}
\newacronym{sm}{SM}{Streaming Multiprocessor}

\begin{abstract}
Current supercomputers often have a heterogeneous architecture using both CPUs and GPUs. At the same time, numerical simulation tasks frequently involve multiphysics scenarios whose components run on different hardware due to multiple reasons, e.g., architectural requirements, pragmatism, etc. This leads naturally to a software design where different simulation modules are mapped to different subsystems of the heterogeneous architecture. We present a detailed performance analysis for such a hybrid four-way coupled simulation of a fully resolved particle-laden flow. The Eulerian representation of the flow utilizes GPUs, while the Lagrangian model for the particles runs on CPUs. First, a roofline model is employed to predict the node level performance and to show that the lattice-Boltzmann-based fluid simulation reaches very good performance on a single GPU. Furthermore, the GPU-GPU communication for a large-scale flow simulation results in only moderate slowdowns due to the efficiency of the CUDA-aware MPI communication, combined with communication hiding techniques. On 1024 A100 GPUs, a parallel efficiency of up to 71\% is achieved. While the flow simulation has good performance characteristics, the integration of the stiff Lagrangian particle system requires frequent CPU-CPU communications that can become a bottleneck. Additionally, special attention is paid to the CPU-GPU communication overhead since this is essential for coupling the particles to the flow simulation. However, thanks to our problem-aware co-partitioning, the CPU-GPU communication overhead is found to be negligible. As a lesson learned from this development, four criteria are postulated that a hybrid implementation must meet for the efficient use of heterogeneous supercomputers. Additionally, an a priori estimate of the speedup for hybrid implementations is suggested.
\end{abstract}

\glsresetall

\maketitle

\section{Introduction}\label{introduction}
The Top500 list reports the most powerful supercomputers worldwide. The number of heterogeneous supercomputers, i.e., systems with additional accelerators such as \glspl{gpu}, in the Top500 list has steadily increased in the last years, accounting for roughly 39\% in June 2024\footnote{\url{www.top500.org/statistics/list/}, Top500 Release: June 2024, category: Accelerator/Co-Processor, accessed on: 18th of October 2024}. Each node of such a heterogeneous supercomputer typically consists of one or many \glspl{cpu} and \glspl{gpu}~\citep[][]{kimSnuCLOpenCLFramework2012}.
Numerical multiphysics simulations are a powerful technique for conducting in-depth investigations of complex physical phenomena by providing detailed data, which is challenging, if not impossible, to collect in experiments.
A significant challenge with these simulations is that they are computationally costly, which is why they are often run on supercomputers.
Especially supercomputers containing \glspl{gpu} have become increasingly popular for numerical simulations in recent years~\citep[][]{shimokawabeStencilFrameworkRealize2017, rohrEnergyEfficientMultiGPUSupercomputer2014, oyarzunPortableImplementationModel2017} as they offer unprecedented computing power.
A common approach in the literature for utilizing such heterogeneous supercomputers for multiphysics simulation is hybrid implementation~\citep[][]{kotsalosDigitalBloodMassively2021, feichtingerDesignPerformanceEvaluation2012, xuDiscreteParticleSimulation2012}, i.e., different simulation modules running on different hardware. Several reasons are predestinating or sometimes even forcing hybrid implementations in the context of multiphysics simulations.
First, the various combined methodologies in such multiphysics simulations may exhibit distinctly contrasting computational properties, e.g., problem sizes, parallel and sequential portions, conditionals, and branching. Therefore, the best-suited hardware architecture can differ between the simulation modules.
Second, if one simulation module dominates the overall run time of the simulation, accelerating only this part using \glspl{gpu} is a straightforward, pragmatic, and development time-saving alternative to porting the whole code base to \gls{gpu}. 
Third, practical limitations can challenge multiphysics simulations: not all coupled software frameworks and modules may support the same hardware, e.g. if parts of the simulation use commercial software.
While hybrid implementations are commonly used in practice for the aforementioned reasons, they introduce inherent challenges in performance and scalability.\newline
A prominent example of multiphysics simulations is coupled fluid-particle simulations with fully resolved particles~\citep[i.e., multiple fluid cells per particle diameter, see][]{rettingerFullyResolvedSimulation2023a}.
Such simulations have been used in the literature, among others, to understand the formation and dynamics of dunes in river beds~\citep[][]{rettingerFullyResolvedSimulations2017, schwarzmeierParticleresolvedSimulationAntidunes2023a, kidanemariamDirectNumericalSimulation2014}, analyze the chimney fluidization in a granular medium~\citep[][]{ngomaTwodimensionalNumericalSimulation2018}, investigate the erosion kinetics of soil under an impinging jet~\citep[][]{benseghierParallelGPUbasedComputational2020} and to analyze mobile sediment beds~\citep[][]{vowinckelFluidParticleInteraction2014, rettingerRheologyMobileSediment2022}.
One encounters some of the previously mentioned reasons for hybrid implementations in the context of coupled fluid-particle simulations with fully resolved particles.
First, when coupling a particle simulation with a fluid simulation using a fully resolved approach, the number of particles is typically at least three orders of magnitude smaller than the number of fluid cells, leading to an imbalance in the workload, where the fluid simulation can overwhelmingly dominate the total run time of the simulation. Therefore, accelerating only the fluid simulation using \glspl{gpu} is a pragmatic approach for this application.
Second, while the fluid simulation employed here uses a structured, cartesian-grid-based methodology and is therefore ideally suited for \gls{gpu} parallelization~\citep[][]{kuznikLBMBasedFlow2010,holzerHighlyEfficientLattice2021}, the situation is more complex for the particle simulation. One extreme is molecular dynamics simulations using millions of particles but relatively simple and uniform particle-particle interactions, which are well suited for \glspl{gpu}~\citep[][]{machadoTinyMDMappingMolecular2021}. The other extreme is particle simulations consisting of few but large particles with complex shapes and sophisticated and diverse particle-particle interactions~\citep[][]{iglbergerMassivelyParallelGranular2010}, assumably better suited for \glspl{cpu}. The particle methodology of this paper lies somewhere in between. While spherical particles are used, the number of particles is comparably small (due to being fully resolved), and the particle-particle interactions are more sophisticated (e.g., lubrication corrections) than the typical molecular dynamics or \gls{dem} simulation on \glspl{gpu} in the literature.
This effect will increase when introducing more complex particle shapes or additional particle-particle interactions (such as cohesion) in the future.\newline
Numerous fluid-particle simulations in the literature use hybrid parallelization in one way or another, of which we present a representative selection in the following.
One example is simulating deformable bodies (using the finite element method) in a fluid, i.e., blood cells in cellular blood flow~\citep[][]{kotsalosDigitalBloodMassively2021}. Here, the focus is more on structural mechanics, as the finite element method dominates the overall run time and is, therefore, the accelerated part of the code. Hybrid approaches have also been used in the literature to couple commercial \gls{cfd} solvers on the \gls{cpu} such as Ansys~\citep[][]{heCPUGPUCrossplatformCoupled2020, sousaniAcceleratedHeatTransfer2019}, or AVL fire~\citep[][]{jajcevicLargescaleCFDSimulations2013} with the \gls{dem} on the \gls{gpu}. Furthermore, simulations have been accelerated using hybrid parallelization to gain a deeper understanding of fluidized beds~\citep[][]{xuDiscreteParticleSimulation2012, norouziNewHybridCPUGPU2017}, or the effects of porosity and tortuosity on soil permeability~\citep[][]{sheikhNumericalInvestigationEffects2015}.
Another variant is to use the \gls{gpu} for both the fluid and the particles, but the \gls{cpu} supports the particle dynamics through collision detection~\citep[][]{juniorFluidSimulationTwoWay2010}.
Furthermore, we point the reader's attention to a related work utilizing a heterogeneous-hybrid approach, i.e., \glspl{cpu} for the particles, while the fluid simulation is distributed on both the \glspl{cpu} and \glspl{gpu}~\citep[][]{feichtingerDesignPerformanceEvaluation2012}.\newline
We chose the \gls{psm} for the coupling between the fluid and solid phase as it has been used successfully in the context of \glspl{gpu}~\citep[][]{benseghierParallelGPUbasedComputational2020, fukumoto2DCoupledFluidparticle2021} due to its \gls{simd} nature.\newline
In this work, we postulate several performance criteria that a hybrid implementation has to meet in our opinion for an efficient use of a heterogeneous supercomputer without misusing resources. First, the overhead introduced by the hybrid implementation (i.e., the \gls{cpu}-\gls{gpu} communication) must be negligible compared to the overall run time. Second, the performance-critical part (in our case the fluid simulation) should show good performance on the \gls{gpu}. Third, the share of the \gls{gpu} run time of the total run time must be sufficiently high to justify using a heterogeneous cluster. Fourth, it has to show a satisfactory weak scaling performance for the efficient use of multiple supercomputer nodes. One of the main contributions is introducing a multi-\gls{cpu}-multi-\gls{gpu} implementation for fully resolved fluid-particle simulations, fulfilling the four criteria stated above.
To evaluate the presented implementation, an in-depth performance analysis on a state-of-the-art heterogeneous supercomputer is provided, comparing two contrasting cases of a fluidized bed simulation, namely in dense and dilute states of the solid granular phase. The paper focuses on the experience with the performance and scalability of the presented hybrid implementation on supercomputers rather than on the exposition of physical results.
Coupled fluid-particle implementations in the literature using a comparable particle methodology still are often limited to \glspl{cpu}~\citep[][]{rettingerFullyResolvedSimulations2017, kidanemariamDirectNumericalSimulation2014, vowinckelFluidParticleInteraction2014}.
With this work, we report and study the experience with the performance and scalability of the presented hybrid implementation on a supercomputer to allow other researchers to make an informed decision if such a `partial' acceleration might be a suitable approach for their codes.\newline
The paper starts in \cref{numerical_methods} with an introduction to the numerical methods, i.e., the \gls{lbm}, the \gls{dem}, and the corresponding coupling. This is followed in \cref{implementation} with a description of the implementation.
\cref{performance} provides a detailed performance analysis. This study includes the \gls{cpu}-\gls{gpu} communication overhead, the run times of the different simulation modules, a roofline model, weak scaling, \change{strong scaling}, and an a priori estimation of the hybrid speedup. \cref{implications} elaborates on the lessons learned and the applicability of the results beyond the present study.
The paper concludes with \cref{conclusion}.

\section{Numerical methods}\label{numerical_methods}
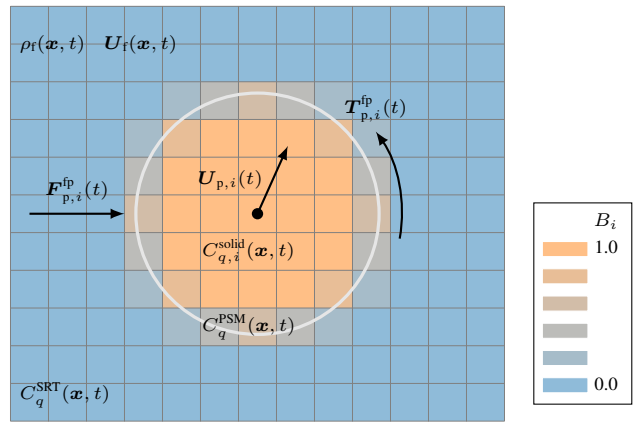
\begin{figure}
  \centering
  \begin{tikzpicture}[scale=1.0]
    \tikzstyle{every node}=[font=\scriptsize]
    
    \definecolor{intermediate1}{HTML}{4C7993}
    \definecolor{intermediate2}{HTML}{797A72}
    \definecolor{intermediate3}{HTML}{A57C50}
    \definecolor{intermediate4}{HTML}{D27D2F}

    \colorlet{blue}{matplotlibBlue!50}
    \colorlet{i1}{intermediate1!50}
    \colorlet{i2}{intermediate2!50}
    \colorlet{i3}{intermediate3!50}
    \colorlet{i4}{intermediate4!50}
    \colorlet{orange}{matplotlibOrange!50}
    
    \fill[blue] (0,0) rectangle (6.5,5.5);
    
    \fill[orange] (2.5,1.5) rectangle ++(1.5,2.5);
    \fill[orange] (2,2) rectangle ++(2.5,1.5);

    \fill[i1] (1.5,1.5) rectangle ++(0.5,0.5);
    \fill[i1] (1.5,3.5) rectangle ++(0.5,0.5);
    \fill[i1] (4.5,1.5) rectangle ++(0.5,0.5);
    \fill[i1] (4.5,3.5) rectangle ++(0.5,0.5);

    \fill[i1] (2,1) rectangle ++(0.5,0.5);
    \fill[i1] (4,1) rectangle ++(0.5,0.5);
    \fill[i1] (2,4) rectangle ++(0.5,0.5);
    \fill[i1] (4,4) rectangle ++(0.5,0.5);

    \fill[i2] (1.5,2) rectangle ++(0.5,0.5);
    \fill[i2] (1.5,3) rectangle ++(0.5,0.5);
    \fill[i2] (4.5,2) rectangle ++(0.5,0.5);
    \fill[i2] (4.5,3) rectangle ++(0.5,0.5);

    \fill[i2] (2.5,1) rectangle ++(0.5,0.5);
    \fill[i2] (3.5,1) rectangle ++(0.5,0.5);
    \fill[i2] (2.5,4) rectangle ++(0.5,0.5);
    \fill[i2] (3.5,4) rectangle ++(0.5,0.5);

    \fill[i3] (1.5,2.5) rectangle ++(0.5,0.5);
    \fill[i3] (4.5,2.5) rectangle ++(0.5,0.5);
    \fill[i3] (3,1) rectangle ++(0.5,0.5);
    \fill[i3] (3,4) rectangle ++(0.5,0.5);

    \fill[i4] (2,1.5) rectangle ++(0.5,0.5);
    \fill[i4] (4,1.5) rectangle ++(0.5,0.5);
    \fill[i4] (2,3.5) rectangle ++(0.5,0.5);
    \fill[i4] (4,3.5) rectangle ++(0.5,0.5);
    
    \draw[step=0.5,gray,very thin] (0,0) grid (6.5,5.5);
    
    \begin{axis}
    [xmin=0,
    xmax=12.925,
    ymin=0,
    ymax=10,
    ticks=none,
    axis lines=none,
    clip=false,
    scale only axis,
    legend pos=south east,
    ]
    \addlegendimage{empty legend}
    \addlegendimage{white,fill=orange,area legend}
    \addlegendimage{white,fill=i4,area legend}
    \addlegendimage{white,fill=i3,area legend}
    \addlegendimage{white,fill=i2,area legend}
    \addlegendimage{white,fill=i1,area legend}
    \addlegendimage{white,fill=blue,area legend}
    \addlegendentry{$B_i$}
    \addlegendentry{1.0}
    \addlegendentry{\phantom{0.8}}
    \addlegendentry{\phantom{0.6}}
    \addlegendentry{\phantom{0.4}}
    \addlegendentry{\phantom{0.2}}
    \addlegendentry{0.0}
    \end{axis}

    \coordinate[
    ] (xpj) at (3.25,2.75);

    \node[fill=black, circle, inner sep=1.5] at (xpj) {};
    \draw[gray!20,very thick] (xpj) circle (1.6);
    \node[right] at (0,5) {$\rho_{\text{f}}(\boldsymbol{x},t) \quad \boldsymbol{U}_{\text{f}}(\boldsymbol{x},t)$\strut};

    \node[right] at (2.4,2.25) {$C_{q,i}^{\text{solid}}(\boldsymbol{x},t)$};
    \node[right] at (2.4,1.25) {$C_q^{\text{PSM}}(\boldsymbol{x},t)$};
    \node[right] at (0,0.35) {$C_q^{\text{SRT}}(\boldsymbol{x},t)$};

    \draw[-latex, thick] (xpj) -- ++(0.4,0.9) node[pos=0.5,left]{$\boldsymbol{U}_{\text{p},i}(t)$};

    \draw[-latex, thick] (0.25,2.75) -- (1.5,2.75) node[pos=0.5,anchor=south]{$\boldsymbol{F}_{\text{p},i}^{\text{fp}}(t)$};

    \draw[-latex,thick] ($(xpj) + (-10:1.9)$) arc (-10:35:1.9);
    \node[above] at ($(xpj) + (35:1.9)$) {$\boldsymbol{T}_{\text{p},i}^{\text{fp}}(t)$};
\end{tikzpicture}
  \caption{
        Two-dimensional sketch of coupled fluid-particle simulations using the \gls{psm}
    }\label{fig:psm_sketch}
\end{figure}

Generally, fully resolved coupled fluid-particle simulations consist of three modules: fluid dynamics, particle physics, and fluid-particle coupling. In this section, we introduce the background of the methods based on the work of~\citet{rettingerComparativeStudyFluidparticle2017, rettingerEfficientFourwayCoupled2022}.
\cref{fig:psm_sketch} illustrates coupled fluid-particle simulations using the \gls{psm}, as it will be explained in the upcoming sections.

\subsection{Lattice Boltzmann method}
We use the \gls{lbm} with the D3Q19 lattice model for the hydrodynamics simulation, an alternative to conventional Navier-Stokes solvers~\citep[][]{krugerLatticeBoltzmannMethod2017}.
We evolve 19 \gls{pdfs} $f_q$ with $q\in \{0,\ldots,18\}$ for every cell of a three-dimensional cartesian lattice. Each $f_q$ is associated with a lattice velocity $\boldsymbol{c}_q$.
The underlying update rule is based on the Boltzmann equation, typically split into the streaming and collision steps.
The cell-local collision relaxes the \gls{pdfs} towards a thermodynamic equilibrium.
The streaming propagates the post-collision \gls{pdfs} $\widetilde{f_q}$ to neighboring cells.
The collision step for the lattice cell $\boldsymbol{x}$ at time step $t$ is defined as
\begin{equation}
  \widetilde{f_q}(\boldsymbol{x},t)=f_q(\boldsymbol{x},t)+C_q(\boldsymbol{x},t)+F_q(\boldsymbol{x},t),
  \label{collision}
\end{equation}
with $C_q$ being the collision operator and $F_q$ the forcing operator. The streaming step
\begin{equation}
  f_q(\boldsymbol{x}+\boldsymbol{c}_q\varDelta t,t+\varDelta t)=\widetilde{f_q}(\boldsymbol{x},t)
\end{equation}
distributes the \gls{pdfs} to neighboring cells. $\varDelta t$ is the time step size, which is typically 1 in the context of the \gls{lbm}. Although recent studies have used more elaborate collision operators in the context of the \gls{psm}~\citep[][]{wangImprovedCouplingTime2018}, the \gls{srt} model (also known as the BGK model~\citep[][]{qianLatticeBGKModels1992}) is still commonly applied in the context of the \gls{psm}, which we will introduce in the following. It relaxes the \gls{pdfs} towards their equilibrium using a single relaxation time $\tau$
\begin{equation}
  C_q^{\text{SRT}}(\boldsymbol{x},t)=\frac{\varDelta t}{\tau}(f_q^{\text{eq}}(\rho_{\text{f}},\boldsymbol{U}_{\text{f}})-f_q(\boldsymbol{x},t)).
  \label{srt_collision}
\end{equation}
The relaxation time $\tau$ is linked to the kinematic fluid viscosity $\nu$ by
\begin{equation}
  \nu=(\tau-\varDelta t/2)c_{\text{s}}^2.
\end{equation}
The equilibrium is defined as
\begin{dmath}
  f_q^{\text{eq}}(\rho_{\text{f}},\boldsymbol{U}_{\text{f}})=w_q\left(\rho_{\text{f}}+\rho_0 \left( \frac{\boldsymbol{c}_q \cdot \boldsymbol{U}_{\text{f}}}{c_{\text{s}}^2} + \frac{{(\boldsymbol{c}_q \cdot \boldsymbol{U}_{\text{f}})}^2}{2c_{\text{s}}^4} \\- \frac{\boldsymbol{U}_{\text{f}} \cdot \boldsymbol{U}_{\text{f}}}{2c_{\text{s}}^2} \right)\right)
\end{dmath}
for incompressible flows~\citep[][]{heLatticeBoltzmannModel1997} with $\rho_0=1$, the lattice weights $w_q$~\citep[][]{qianLatticeBGKModels1992}, and the lattice speed of sound $c_{\text{s}}=1/\sqrt{3}$. The cell-local quantities
\begin{equation}
  \rho_{\text{f}}(\boldsymbol{x},t)=\sum_q f_q(\boldsymbol{x},t),
\end{equation}
\begin{equation}
  \boldsymbol{U}_{\text{f}}(\boldsymbol{x},t)=\frac{1}{\rho_0}\sum_q f_q(\boldsymbol{x},t)\boldsymbol{c}_q+\frac{\varDelta t}{2\rho_0}\boldsymbol{f}^{\text{ext}}
\end{equation}
are calculated based on the moments of the \gls{pdfs}.
The forcing operator
\begin{equation}
  F_q(\boldsymbol{x},t)=\varDelta t w_q \left[\frac{\boldsymbol{c}_q-\boldsymbol{U}_{\text{f}}}{c_{\text{s}}^2} + \frac{\boldsymbol{c}_q\cdot\boldsymbol{U}_{\text{f}}}{c_{\text{s}}^4} \cdot\boldsymbol{c}_q\right]\cdot\boldsymbol{f}^{\text{ext}}
\end{equation}
can incorporate external forces using the constant force density $\boldsymbol{f}^{\text{ext}}$~\citep[][]{laddLatticeBoltzmannSimulationsParticleFluid2001}.

\subsection{Particle dynamics}
The behavior of the particles is modeled using the \gls{dem}~\citep[][]{cundallDiscreteNumericalModel1979}. The total force $\boldsymbol{F}_{\text{p},i}$ acting on a particle $i$ consists of the following modules
\begin{equation}
  \boldsymbol{F}_{\text{p},i}=\boldsymbol{F}_{\text{p},i}^{\text{col}}+\boldsymbol{F}_{\text{p},i}^{\text{hyd}}+\boldsymbol{F}_{\text{p},i}^{\text{ext}}.
\end{equation}
In addition to the hydrodynamic force $\boldsymbol{F}_{\text{p},i}^{\text{hyd}}$ and external forces $\boldsymbol{F}_{\text{p},i}^{\text{ext}}$ (e.g., gravity), the particle interactions exert forces $\boldsymbol{F}_{\text{p},i}^{\text{col}}$ on each other due to collisions. The equations of motion have to be integrated to simulate the particle movements.

\subsubsection{Particle interactions using the discrete element method}
The collision between particle $i$ and $j$ is modeled using a linear spring-dashpot model. The collision force $\boldsymbol{F}_{\text{p},i}^{\text{col}}$ and torque $\boldsymbol{T}_{\text{p},i}^{\text{col}}$ on particle $i$ are computed as
\begin{equation}
  \boldsymbol{F}_{\text{p},i}^{\text{col}}=\sum_{j,j \neq i}(\boldsymbol{F}_{ij,\text{n}}^{\text{col}}+\boldsymbol{F}_{ij,\text{t}}^{\text{col}}),
\end{equation}
\begin{equation}
  \boldsymbol{T}_{\text{p},i}^{\text{col}}=\sum_{j,j \neq i}(\boldsymbol{x}_{ij}^{\text{cp}}-\boldsymbol{x}_{\text{p},i})\times \boldsymbol{F}_{ij,\text{t}}^{\text{col}},
\end{equation}
where the normal part of the collision force $\boldsymbol{F}_{ij,\text{n}}^{\text{col}}$ acting on particle $i$ with position $\boldsymbol{x}_{\text{p},i}$ is computed as
\begin{equation}
  \boldsymbol{F}_{ij,\text{n}}^{\text{col}}=-k_{\text{n}}\delta_{ij,\text{n}}\boldsymbol{n}_{ij}-d_{\text{n}}\boldsymbol{U}_{ij,\text{n}}^{\text{cp}}.
\end{equation}
Here, $k_{\text{n}}$ and $d_{\text{n}}$ are the normal stiffness and damping coefficients, $\boldsymbol{n}_{ij}$ the normal vector, $\delta_{ij,\text{n}}$ is the penetration depth and $\boldsymbol{U}_{ij,\text{n}}^{\text{cp}}$ is the normal component of the relative velocity of the surface of the particle at the contact point $\boldsymbol{x}_{ij}^{\text{cp}}$.
The tangential part of the collision force
\begin{equation}
  \boldsymbol{F}_{ij,\text{t}}^{\text{col}}=-k_{\text{t}}\boldsymbol{\delta}_{ij,\text{t}}-d_{\text{t}}\boldsymbol{U}_{ij,\text{t}}^{\text{cp}}
\end{equation}
uses the tangential stiffness and damping coefficients $k_{\text{t}}$ and $d_{\text{t}}$ and $\boldsymbol{U}_{ij,\text{t}}^{\text{cp}}$ is the tangential component of the relative velocity of the surface of the particle at the contact point.
\begin{equation}
  \boldsymbol{\delta}_{ij,\text{t}}=\int_{t_i}^{t} \boldsymbol{U}_{ij,\text{t}}^{\text{cp}}(t')\text{d}t'
  \label{history_information}
\end{equation}
is the accumulated relative tangential motion between two particles where $t_i$ is the time step of the impact. For more details, see \citet{rettingerEfficientFourwayCoupled2022}.

%
%
%


\subsubsection{Integration of the particle properties}\label{integration}
We update the particle's position and velocity by solving the Newton-Euler equations of motion using the Velocity Verlet integrator:
\begin{equation}
  \boldsymbol{x}_{\text{p},i}(t+\varDelta t_{\text{p}})=\boldsymbol{x}_{\text{p},i}(t)+\varDelta t_{\text{p}}\boldsymbol{U}_{\text{p},i}(t)+\frac{\varDelta t_{\text{p}}^2}{2m_{\text{p},i}}\boldsymbol{F}_{\text{p},i}(t),
  \label{pre_force_integration}
\end{equation}
\begin{dmath}
  \boldsymbol{U}_{\text{p},i}(t+\varDelta t_{\text{p}})=\boldsymbol{U}_{\text{p},i}(t)+\frac{\varDelta t_{\text{p}}}{2m_{\text{p},i}}(\boldsymbol{F}_{\text{p},i}(t)\\+\boldsymbol{F}_{\text{p},i}(t + \varDelta t_{\text{p}})),
  \label{post_force_integration}
\end{dmath}
where $m_{\text{p},i}$ is the mass of the particle $i$. $\boldsymbol{x}_{\text{p},i}(t+\varDelta t_{\text{p}})$ is computed at the beginning of each particle time step using the old force. Then, the new force $\boldsymbol{F}_{\text{p},i}(t + \varDelta t_{\text{p}})$ is computed using the updated position. At the end of the time step, the particle velocity $\boldsymbol{U}_{\text{p},i}(t+\varDelta t_{\text{p}})$ is computed using the updated force. Updating the angular velocity is done analogously.

\subsection{Fully resolved fluid-particle coupling method}\label{psm}
The task of the coupling is to perform momentum exchange between the fluid and the solid phase. 
We use the \gls{psm} for the fully resolved fluid-particle coupling~\citep[][]{nobleLatticeBoltzmannMethodPartially1998}. It modifies the \gls{lbm} collision step from \cref{collision} by introducing the solid volume fraction $B(\boldsymbol{x},t)$ resulting in
\begin{equation}
  \widetilde{f_q}(\boldsymbol{x},t)=f_q(\boldsymbol{x},t)+C_q^{\text{PSM}}(\boldsymbol{x},t)+(1-B(\boldsymbol{x},t))F_q(\boldsymbol{x},t),
  \label{psm_equation}
\end{equation}
where $B(\boldsymbol{x},t)$ is the fraction of the fluid cell $\boldsymbol{x}$ being (partly) covered by one or more particles. \cref{particle_mapping} explains this solid volume fraction computation in detail. The modified collision operator $C_q^{\text{PSM}}$ used in \cref{psm_equation} is defined as
\begin{dmath}
  C_q^{\text{PSM}}(\boldsymbol{x},t)=(1-B(\boldsymbol{x},t))C_q^{\text{SRT}}(\boldsymbol{x},t)\\+\sum_{I} B_i(\boldsymbol{x},t)C_{q,i}^{\text{solid}}(\boldsymbol{x},t).
\end{dmath}
$C_q^{\text{SRT}}$ is the \gls{lbm} collision operator described in \cref{srt_collision}. $B(\boldsymbol{x},t)$ is the sum over the individual overlap fractions $B_i(\boldsymbol{x},t)$ of all particles $I$. If $B(\boldsymbol{x},t) > 1$, it is normalized to 1. This situation can occur if colliding particles are allowed to overlap during contact. Then a single fluid cell can even be entirely covered by two particles, i.e., $B(\boldsymbol{x},t) = 2$.
The solid collision operator
\begin{dmath}
  C_{q,i}^{\text{solid}}(\boldsymbol{x},t)=[f_{\bar{q}}(\boldsymbol{x},t)-f_{\bar{q}}^{\text{eq}}(\rho_{\text{f}},\boldsymbol{U}_{\text{f}})]-[f_q(\boldsymbol{x},t)\\-f_q^{\text{eq}}(\rho_{\text{f}},\boldsymbol{U}_{\text{p},i}(\boldsymbol{x},t))]
\end{dmath}
acts when particles intersect with a cell.
There exist different variants of the solid collision operator.
$f_{\bar{q}}$ corresponds to the inverse lattice velocity of $f_q$.\newline
$\boldsymbol{U}_{\text{p},i}(\boldsymbol{x},t)$ is the velocity of particle $i$ evaluated at the cell center $\boldsymbol{x}$ and is computed as
\begin{equation}
  \boldsymbol{U}_{\text{p},i}(\boldsymbol{x}_i,t)=\boldsymbol{U}_{\text{p},i}(t)+\boldsymbol{\Omega}_{\text{p},i}(t) \times (\boldsymbol{x}_i-\boldsymbol{x}_{\text{p},i}(t)),
  \label{vel_at_cell_center}
\end{equation}
with the translational particle velocity $\boldsymbol{U}_{\text{p},i}(t)$, the rotational particle velocity ${\boldsymbol{\Omega}}_{\text{p},i}(t)$ and the particle center of gravity $\boldsymbol{x}_{\text{p},i}(t)$. $\boldsymbol{x}_i$ are the cell centers of all cells intersecting with the particle.\newline
So far, we have only considered the influence of the particles on the fluid. However, the fluid also influences the particles through hydrodynamic forces. We compute the force $\boldsymbol{F}_{\text{p},i}^{\text{fp}}(t)$ and torque $\boldsymbol{T}_{\text{p},i}^{\text{fp}}(t)$ exerted by the fluid on particle $i$ as
\begin{equation}
  \boldsymbol{F}_{\text{p},i}^{\text{fp}}(t)=\frac{{(\varDelta x)}^3}{\varDelta t}\sum_{\boldsymbol{x_i}}[B_i(\boldsymbol{x}_i,t)\sum_q (C_{q,i}^{\text{solid}}(\boldsymbol{x}_i,t)\boldsymbol{c}_{\bar{q}})],
  \label{hydrodynamic_force}
\end{equation}
\begin{dmath}
  \boldsymbol{T}_{\text{p},i}^{\text{fp}}(t)=\frac{{(\varDelta x)}^3}{\varDelta t}\sum_{\boldsymbol{x_i}}[B_i(\boldsymbol{x}_i,t)(\boldsymbol{x}_i-\boldsymbol{x}_{\text{p},i})\\\times\sum_q (C_{q,i}^{\text{solid}}(\boldsymbol{x}_i,t)\boldsymbol{c}_{\bar{q}})].\label{hydrodynamic_torque}
\end{dmath}

\subsubsection{Lubrication correction}\label{lubrication_correction}
The lubrication force and torque act on two particles approaching each other.
The two particles squeeze out the fluid inside the gap, which exerts a force in the opposite direction of the relative motion.
However, this effect would only be covered correctly by the fluid-particle coupling for a very fine grid resolution which is computationally too expensive.
As the lubrication force has a significant influence, we compute lubrication correction force terms to compensate for the inability of the coupling method to represent these forces correctly.
We compute lubrication correction terms due to normal- and tangential translations and rotations.
Therefore, the total hydrodynamic force $\boldsymbol{F}_{\text{p},i}^{\text{hyd}}$ and torque $\boldsymbol{T}_{\text{p},i}^{\text{hyd}}$ is a sum of the force from the fully resolved fluid-particle coupling method (\cref{psm}), and the lubrication correction
\begin{equation}
  \boldsymbol{F}_{\text{p},i}^{\text{hyd}}=\boldsymbol{F}_{\text{p},i}^{\text{fp}}+\boldsymbol{F}_{\text{p},i}^{\text{lub,cor}},
\end{equation}
\begin{equation}
  \boldsymbol{T}_{\text{p},i}^{\text{hyd}}=\boldsymbol{T}_{\text{p},i}^{\text{fp}}+\boldsymbol{T}_{\text{p},i}^{\text{lub,cor}}.
\end{equation}
For details on how to compute the lubrication correction, see~\citet{rettingerEfficientFourwayCoupled2022}.

%
%
%
%
%
%
%

\subsubsection{Particle mapping}\label{particle_mapping}
A coupled fluid-particle simulation using the \gls{psm} requires the computation of the solid volume fraction $B_i(\boldsymbol{x},t)$ (\cref{psm}), i.e., the fraction of a fluid cell $\boldsymbol{x}$ being (partly) covered by a particle $i$. We restrict ourselves to spherical particles. \citet{jonesFastComputationAccurate2017} tackle the problem that, in general, no unique analytical solution exists to compute this overlapping fraction for spheres and cells/cubes. They propose a linear approximation derived from the analytical solution for a specific cell orientation relative to the particle surface. Grid cells with the dimensionless edge size 1 (as it is typically the case for the \gls{lbm}) are assumed in the following.
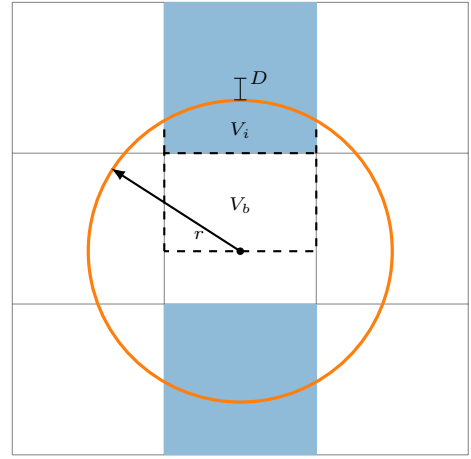
\begin{figure}
  \centering
  \begin{tikzpicture}[scale=1.0]
    \tikzstyle{every node}=[font=\scriptsize]
    \colorlet{lightBlue}{matplotlibBlue!50}
    
    \draw[step=2.0,gray,very thin] (0,0) grid (6,6);

    \fill[lightBlue] (2,4) rectangle ++(2,2);
    \fill[lightBlue] (2,0) rectangle ++(2,2);
    
    \begin{axis}
    [xmin=0,
    xmax=12.925,
    ymin=0,
    ymax=10,
    ticks=none,
    axis lines=none,
    clip=false,
    scale only axis
    ]
    \end{axis}
    
    \coordinate[] (xpj) at (3,2.7);
    \node[fill=black, circle, inner sep=1.0] at (xpj) {};
    \draw[matplotlibOrange,very thick] (xpj) circle (2.0);

    \draw[thick, dashed] (2,2.7) rectangle (4,4);
    \draw[thick, dashed] (2,4)--(2,4.4);
    \draw[thick, dashed] (4,4)--(4,4.4);
    \node[] at (3,3.3) {$V_b$\strut};
    \node[] at (3,4.3) {$V_i$\strut};

    \draw[-latex, thick] (xpj) -- ++(-1.7,1.1) node[pos=0.2,left]{$r$};

    \draw[black, |-|,  >={Latex[scale=0.75]}] (3,4.7) -- (3,5) node[right]{$D$};
\end{tikzpicture}
  \caption{
        The linear approximation yields the analytical solution for the blue cells. The particle is represented by the orange circle. Note that the grid is coarsened for better clarity.
    }\label{fig:linear_approximation}
\end{figure}
In \cref{fig:linear_approximation}, the overlap fraction $\epsilon$ between the upper blue lattice cell and the orange particle is computed as
\begin{equation}
  \epsilon=V_i=V_a-V_b=V_a-(D+r-1/2),
\end{equation}
where $V_a$ is the union of $V_i$ and $V_b$. $D$ is the distance from the cell center to the sphere surface (negative if the cell center lies inside the sphere). There is a cell-particle overlap if $\epsilon \in \interval[open left]{0}{1}$. We can reformulate this as
\begin{equation}
  \epsilon=V_a-(D+r-1/2)=-D+f(r),
\end{equation}
where $f(r)=V_a-r+1/2$ only depends on the particle radius $r$ and therefore is constant for each particle respectively.
$V_a$ is computed as
\begin{dmath}
  V_a=\int_{-1/2}^{1/2}\int_{-1/2}^{1/2} \sqrt{r^2-x^2-y^2} \text{d}x \text{d}y= (1/12-r^2)\tan^{-1} (\frac{\frac{1}{2}\sqrt{{r^2}-1/2}}{1/2-r^2})+\frac{1}{3}\sqrt{{r^2}-1/2}\\
  + (r^2- 1/12)\tan^{-1} (\frac{1/2}{\sqrt{{r^2}-1/2}})\\-\frac{4}{3}r^3\tan^{-1} (\frac{1/4}{r\sqrt{{r^2}-1/2}}).
\end{dmath}
This approximation yields accurate results also for arbitrary cell orientations and is more computationally efficient than contemporary techniques like sub-division sampling \citep[][]{jonesFastComputationAccurate2017}.

\section{Implementation}\label{implementation}
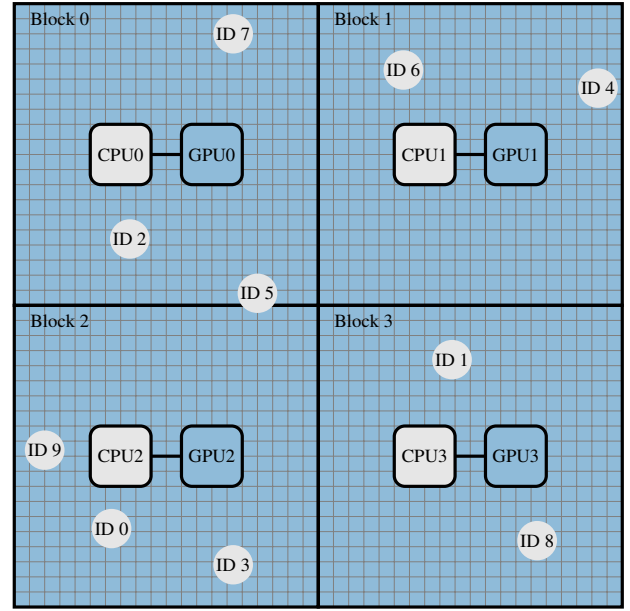
\begin{figure}
  \centering
  \begin{tikzpicture}[scale=0.8]
    \tikzstyle{every node}=[font=\scriptsize]

    \colorlet{blue}{matplotlibBlue!50}
    
    \fill[blue] (0,0) rectangle (10.0,10.0);
    
    \draw[step=0.25,gray,very thin] (0,0) grid (10.0,10.0);
    \draw[black,very thick] (0,0) rectangle (5,5);
    \draw[black,very thick] (5,0) rectangle (10,5);
    \draw[black,very thick] (0,5) rectangle (5,10);
    \draw[black,very thick] (5,5) rectangle (10,10);

    \node[black] at (0.75,9.75) {Block 0};
    \node[black] at (5.75,9.75) {Block 1};
    \node[black] at (0.75,4.75) {Block 2};
    \node[black] at (5.75,4.75) {Block 3};

    \draw[gray!20, fill=gray!20, very thick] (1.6,1.3) circle (0.3) node [black] {ID 0};
    \draw[gray!20, fill=gray!20, very thick] (7.2,4.1) circle (0.3) node [black] {ID 1};
    \draw[gray!20, fill=gray!20, very thick] (1.9,6.1) circle (0.3) node [black] {ID 2};
    \draw[gray!20, fill=gray!20, very thick] (3.6,0.7) circle (0.3) node [black] {ID 3};
    \draw[gray!20, fill=gray!20, very thick] (9.6,8.6) circle (0.3) node [black] {ID 4};
    \draw[gray!20, fill=gray!20, very thick] (4,5.2) circle (0.3) node [black] {ID 5};
    \draw[gray!20, fill=gray!20, very thick] (6.4,8.9) circle (0.3) node [black] {ID 6};
    \draw[gray!20, fill=gray!20, very thick] (3.6,9.5) circle (0.3) node [black] {ID 7};
    \draw[gray!20, fill=gray!20, very thick] (8.6,1.1) circle (0.3) node [black] {ID 8};
    \draw[gray!20, fill=gray!20, very thick] (0.5,2.6) circle (0.3) node [black] {ID 9};

    \draw[black, fill=gray!20, very thick, rounded corners] (1.25,7) rectangle (2.25,8) node [black, pos=.5] {CPU0};
    \draw[black, fill=blue, very thick, rounded corners] (2.75,7) rectangle (3.75,8) node [black, pos=.5] {GPU0};
    \draw[black, fill=gray!20, very thick, rounded corners] (6.25,7) rectangle (7.25,8) node [black, pos=.5] {CPU1};
    \draw[black, fill=blue, very thick, rounded corners] (7.75,7) rectangle (8.75,8)  node [black, pos=.5] {GPU1};
    \draw[black, fill=gray!20, very thick, rounded corners] (1.25,2) rectangle (2.25,3) node [black, pos=.5] {CPU2};
    \draw[black, fill=blue, very thick, rounded corners] (2.75,2) rectangle (3.75,3) node [black, pos=.5] {GPU2};
    \draw[black, fill=gray!20, very thick, rounded corners] (6.25,2) rectangle (7.25,3) node [black, pos=.5] {CPU3};
    \draw[black, fill=blue, very thick, rounded corners] (7.75,2) rectangle (8.75,3) node [black, pos=.5] {GPU3};

    \draw [black, very thick] (2.25,7.5) -- (2.75,7.5);
    \draw [black, very thick] (7.25,7.5) -- (7.75,7.5);
    \draw [black, very thick] (2.25,2.5) -- (2.75,2.5);
    \draw [black, very thick] (7.25,2.5) -- (7.75,2.5);
\end{tikzpicture}
  \caption{
        Partitioning of a 2D simulation domain into four blocks. The circles with ID 0 to ID 9 indicate the particles, and the blue cells are the fluid. One \gls{gpu}x for updating the fluid cells is assigned to each block x, and the corresponding \gls{cpu}x cores are responsible for the particle dynamics. \gls{cpu}x represents the \gls{cpu} cores having a direct connection/affinity to \gls{gpu}x. MPI rank x is assigned to \gls{cpu}x, distributes the particle computations among \gls{cpu}x using OpenMP, and uses \gls{gpu}x for the fluid dynamics.
    }\label{fig:domain_partitioning}
\end{figure}

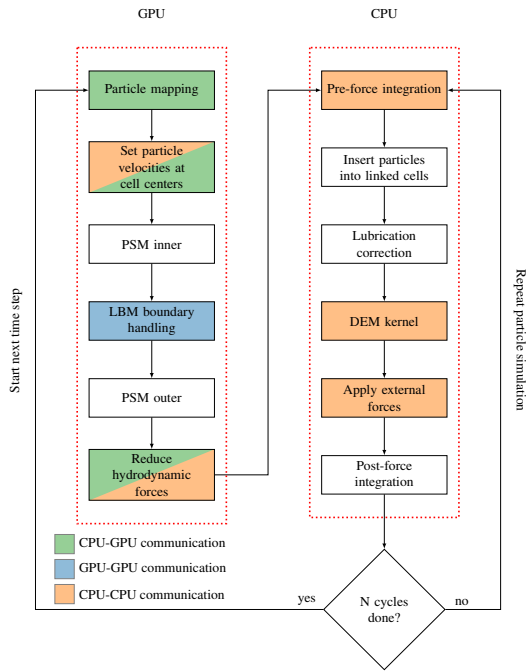
\begin{figure}
  \centering
  \resizebox{7cm}{!}{
\begin{tikzpicture}[scale=1.0, node distance=2cm]
    \node (mapping) [hybridKernel] {Particle mapping};
    \node (setU) [syncKernel, below of=mapping, draw=none] {\textcolor{matplotlibOrange!50}{Set particle velocities at cell centers}};
    \fill[minimum width=3cm, matplotlibGreen!50] (setU.south west) -- (setU.south east) -- (setU.north east);
    \node (setU2) [hybridKernel, below of=mapping, fill=none]{Set particle velocities at cell centers};
    \node (psmInner) [kernel, below of=setU] {PSM inner};
    \node (bh) [syncSweep, below of=psmInner] {LBM boundary handling};
    \node (psmOuter) [kernel, below of=bh] {PSM outer};
    \node (redF) [hybridKernel, below of=psmOuter, draw=none] {\textcolor{matplotlibGreen!50}{Reduce hydrodynamic forces}};
    \fill[minimum width=3cm, matplotlibOrange!50] (redF.south west) -- (redF.south east) -- (redF.north east);
    \node (redF2) [syncKernel, below of=psmOuter, fill=none]{Reduce hydrodynamic forces};

    \node (preForceInt) [syncKernel, right of=mapping, xshift=4cm] {Pre-force integration};
    \node (linkedCells) [kernel, below of=preForceInt] {Insert particles into linked cells};
    \node (lubCorr) [kernel, below of=linkedCells] {Lubrication correction};
    \node (dem) [syncKernel, below of=lubCorr] {DEM kernel};
    \node (externalF) [syncKernel, below of=dem] {Apply external forces};
    \node (postForceInt) [kernel, below of=externalF] {Post-force integration};

    \node (decision) [decision, below of=postForceInt, yshift=-1.5cm, text width=2cm] {N cycles done?};

    \draw [-latex] (mapping) -- (setU);
    \draw [-latex] (setU) -- (psmInner);
    \draw [-latex] (psmInner) -- (bh);
    \draw [-latex] (bh) -- (psmOuter);
    \draw [-latex] (psmOuter) -- (redF);
    \draw [-latex] (redF) - ++(3,0) -- ++(3,10) -- (preForceInt);

    \draw [-latex] (preForceInt) -- (linkedCells);
    \draw [-latex] (linkedCells) -- (lubCorr);
    \draw [-latex] (lubCorr) -- (dem);
    \draw [-latex] (dem) -- (externalF);
    \draw [-latex] (externalF) -- (postForceInt);
    \draw [-latex] (postForceInt) -- (decision);
    \draw [-latex] (decision)node[anchor=south, xshift=2cm] {no} - ++(3,0) -- ++(3,13.5) -- (preForceInt);
    \draw [-latex] (decision)node[anchor=south, xshift=-2cm] {yes} - ++(-9,0) -- ++(-9,13.5) -- (mapping);

    \node (nextStep) [rotate=90] at (-3.5,-6.5) {Start next time step};
    \node (nextCycle) [rotate=-90] at (9.5,-6.5) {Repeat particle simulation};

    \draw[red,very thick,dotted] ($(mapping.north west)+(-0.3,0.6)$)  rectangle ($(redF.south east)+(0.3,-0.6)$);
    \draw[red,very thick,dotted] ($(preForceInt.north west)+(-0.3,0.6)$)  rectangle ($(postForceInt.south east)+(0.3,-0.6)$);
    \node[] at (0,2) {GPU};
    \node[] at (6,2) {CPU};

    \draw[gray, very thin, fill=matplotlibGreen!50] (-2.5,-12.05) rectangle ++(0.5,0.5);
    \node[right] at (-2,-11.8) {\gls{cpu}-\gls{gpu} communication\strut};
    \draw[gray, very thin, fill=matplotlibBlue!50] (-2.5,-12.7) rectangle ++(0.5,0.5);
    \node[right] at (-2,-12.45) {\gls{gpu}-\gls{gpu} communication\strut};
    \draw[gray, very thin, fill=matplotlibOrange!50] (-2.5,-13.35) rectangle ++(0.5,0.5);
    \node[right] at (-2,-13.15) {\gls{cpu}-\gls{cpu} communication\strut};
\end{tikzpicture}
}
  \caption{
        Flowchart of our hybrid \gls{cpu}-\gls{gpu} implementation from the perspective of a \gls{cpu} and \gls{gpu} responsible for the same block. The color coding indicates the communication types required within each step.
    }\label{fig:implementation}
\end{figure}

We implemented our hybrid coupled fluid-particle simulation within the massively parallel multiphysics framework \walberla~\citep[][]{bauerWaLBerlaBlockstructuredHighperformance2021a} (\url{https://www.walberla.net/}).\ \walberla~supports highly efficient and scalable \gls{lbm} simulations on both \glspl{cpu} and \glspl{gpu}~\citep[][]{holzerHighlyEfficientLattice2021}.
The MesaPD module~\citep[][]{eiblModularExtensibleSoftware2019} enables \walberla~to perform particle simulations on \glspl{cpu} using the \gls{dem}.
Large-scale simulations require the usage of numerous nodes of a supercomputer. Each node of a heterogeneous supercomputer typically consists of one or many \glspl{cpu} and \glspl{gpu}. Several \gls{cpu} cores belong to one \gls{gpu}, i.e., they have a direct connection. We divide the simulation domain into multiple blocks and exclusively assign each block to a \gls{gpu}. Figure~\ref{fig:domain_partitioning} illustrates the domain partitioning and the respective hardware assignment.
We want to highlight that the particle simulation is not a standalone framework, but for the particles and the fluid that are physically close to each other (i.e., in the same block), one MPI process is responsible for both the particle and fluid dynamics and the \gls{cpu} and \gls{gpu} have a direct connection/affinity.
The \gls{cpu} cores belonging to the respective \gls{gpu} are responsible for updating the particles whose center of mass lies inside that block (local particles). In Figure~\ref{fig:domain_partitioning}, \gls{cpu}0 is responsible for the particles with ID 2, 5, and 7.
Additionally, particles can overlap with a given block whose center of mass lies in another block (ghost particles). In Figure~\ref{fig:domain_partitioning}, the particle with ID 5 is local for block 0 and ghost for block 2. This overlapping causes the need for communication between the \glspl{cpu}. The particle computations within a block are parallelized among the \gls{cpu} cores using OpenMP.~The communication between neighboring blocks is implemented using the CUDA-Aware Message Passing Interface. On clusters with multiple \glspl{gpu} sharing a node and NVLinks between the \glspl{gpu}, NVIDIA GPUDirect is used for direct GPU-GPU MPI communications between the \glspl{gpu} on the node.
\cref{fig:implementation} illustrates the different modules of the simulation, on which hardware they are running, the workflow, and the necessary communication steps. We will explain the figure in detail in the following sections. Generally speaking, the \gls{gpu} is responsible for all operations on fluid cells (i.e., the \gls{lbm} and the coupling), whereas the \gls{cpu}  performs all computations on particles. The associated data structures are consequently located in the respective memories (fluid cells in \gls{gpu} memory, particles in \gls{cpu} memory).

\subsection{Fluid dynamics and coupling on the GPU}
Performing an \gls{lbm} update step requires the communication of boundary cells between neighboring \glspl{gpu}. However, the first three kernels do not need neighboring information. Therefore, the communication is hidden behind those kernels by starting a non-blocking send before the particle mapping. A time step begins with the coupling from the particles to the fluid.

\subsubsection{Coupling from the particles to the fluid}
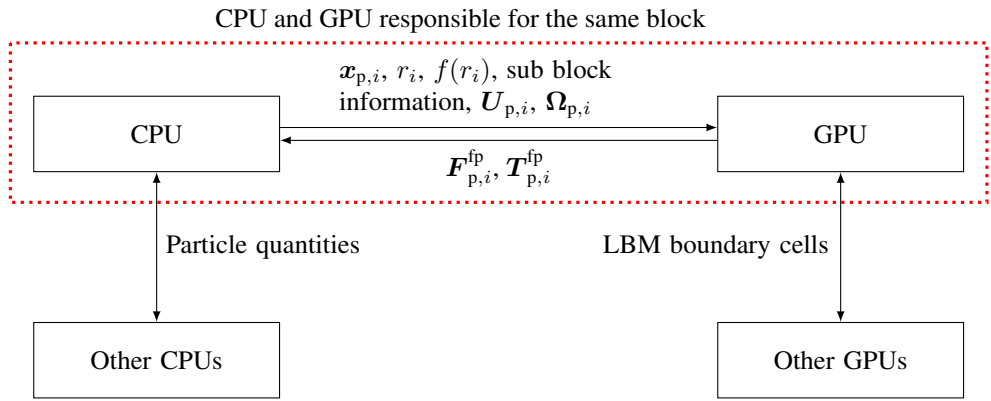
\begin{figure*}
  \centering
\begin{tikzpicture}[scale=1.0]
    \node (cpu) [kernel] {CPU};
    \node (cpus) [kernel, below of=cpu, yshift=-2cm] {Other CPUs};
    \node (gpu) [kernel, right of=cpu, xshift=8cm] {GPU};
    \node (gpus) [kernel, below of=gpu, yshift=-2cm] {Other GPUs};

    \draw[red,very thick,dotted] ($(cpu.north west)+(-0.3,0.7)$)  rectangle ($(gpu.south east)+(0.3,-0.4)$);
    \node[] at (4,1.5) {CPU and GPU responsible for the same block};

    \draw [-latex] ([yshift=-12pt]cpu.north east) -- node (nextStep) [above, text width=5cm, xshift=0.4cm] {$\boldsymbol{x}_{\text{p},i}$, $r_i$, $f(r_i)$, sub block information, $\boldsymbol{U}_{\text{p},i}$, ${\boldsymbol{\Omega}}_{\text{p},i}$} ([yshift=-12pt]gpu.north west);
    \draw [-latex] ([yshift=12pt] gpu.south west) -- node (nextStep) [below] {$\boldsymbol{F}_{\text{p},i}^{\text{fp}}$, $\boldsymbol{T}_{\text{p},i}^{\text{fp}}$} ([yshift=12pt] cpu.south east);
    \draw [latex-latex] (cpu) -- node (nextStep) [right, text width=3cm] {Particle quantities} (cpus); 
    \draw [latex-latex] (gpu) -- node (nextStep) [left, text width=3cm] {LBM boundary cells} (gpus);
\end{tikzpicture}
  \caption{
        Overview of the different communication steps from the perspective of a \gls{cpu} and \gls{gpu} responsible for the same block
    }\label{fig:compute_setup}
\end{figure*}

For the particle mapping, the \gls{gpu} has to check overlaps for all cell-particle combinations, even though there is no overlap for most cell-particle combinations. This check quickly becomes very computationally expensive. Therefore, we reduce the computational effort by dividing each block into $k$ sub-blocks in each dimension.
The \gls{cpu} inserts every particle into all sub-blocks that overlap with this particle, similar to the Linked Cell Method. Using sub-blocks allows the \gls{gpu} to consider only a tiny subset of particles when computing the overlaps for a particular grid cell, namely the particles overlapping with the sub-block the cell is located in.
Our coupled fluid-particle simulation requires the communication of various data. \cref{fig:compute_setup} gives an overview of the different types of communication from the perspective of a \gls{cpu} and \gls{gpu} responsible for the same block. We will explain the communication steps in the following.
For every particle $i$, the position $\boldsymbol{x}_{\text{p},i}$, radius $r_i$, and $f(r_i)$ (\cref{particle_mapping}) are transferred from the \gls{cpu} to the \gls{gpu} via the PCIe. In addition, the number of overlapping particles per sub-block and the corresponding particle IDs are transferred from the \gls{cpu} to the \gls{gpu}.
Then, the \gls{gpu} performs the particle mapping. For more details, see the description of the solid volume fraction computation (i.e., the particle mapping) in \cref{particle_mapping}.
In our simulation, a maximum of two particles can overlap with a given grid cell due to the geometrically resolved spherical particles and appropriate \gls{dem} parameters allowing only a small particle-particle penetration. Therefore, the grid that we use to store $B_i(\boldsymbol{x},t)$, can store two fraction values per grid cell.
Next, the linear and angular velocities $\boldsymbol{U}_{\text{p},i}$ and ${\boldsymbol{\Omega}}_{\text{p},i}$ of the particles must be synchronized between the \glspl{cpu} such that every \gls{cpu} has not only the correct velocities for its local particles but also for the ghost particles. Next, those velocities are transferred from the \gls{cpu} to the \gls{gpu} so that the \gls{gpu} can compute the velocities of the overlapping particles at the cell center for every cell (\cref{vel_at_cell_center}).

\subsubsection{Fluid simulation}\label{fluid_simulation}
Next, the \gls{psm} inner kernel is performed. The term `inner' indicates that this kernel updates all cells except the outermost layer of cells. Skipping the outermost layer ensures that this routine can be called without waiting for the previously started \gls{gpu}-\gls{gpu} communication to finish.
The \gls{psm} kernel creates the highest workload of the entire simulation. It is, therefore, performance-critical. We use the code generation framework lbmpy~\citep[][]{bauerLbmpyAutomaticCode2021, hennigAdvancedAutomaticCode2023} to obtain highly efficient and scalable \gls{lbm} CUDA kernels.\ lbmpy allows the formulation of arbitrary \gls{lbm} methods (such as the \gls{psm}) as a symbolic representation and generates optimized and parallel compute kernels. We integrate those generated compute kernels within the simulation in \walberla.
Next, we wait for the non-blocking \gls{gpu}-\gls{gpu} communication started at the beginning of the time step to finish. Depending on the available hardware, this may return instantly if the previous computations completely hide the communication.
The next step is the \gls{lbm} boundary handling, which enforces boundary conditions to the fluid simulation by correctly updating the fluid cells at the domain's boundary.
Since the neighboring values are now available, we then update the outermost layer of cells in the \gls{psm} outer kernel.
The last step on the \gls{gpu} is the coupling from the fluid to the particles.

\subsubsection{Coupling from the fluid to the particles}
Finally, the \gls{gpu} reduces the forces and torques exerted by the fluid on the particles $\boldsymbol{F}_{\text{p},i}^{\text{fp}}$ and $\boldsymbol{T}_{\text{p},i}^{\text{fp}}$ (\cref{hydrodynamic_force,hydrodynamic_torque}).
Then, $\boldsymbol{F}_{\text{p},i}^{\text{fp}}$ and $\boldsymbol{T}_{\text{p},i}^{\text{fp}}$ are transfered from the \gls{gpu} to the \gls{cpu} to be available for the upcoming \gls{dem} simulation on the \gls{cpu}. 
A single particle may overlap with cells located on multiple blocks. Thus, multiple \glspl{gpu} may have computed $\boldsymbol{F}_{\text{p},i}^{\text{fp}}$ and $\boldsymbol{T}_{\text{p},i}^{\text{fp}}$ for the same particle $i$. Therefore, the corresponding \glspl{cpu} have to reduce these force and torque contributions exerted by the coupling on the particles into a single variable (\gls{cpu}-\gls{cpu} communication).
The time loop continues with the particle dynamics on the \gls{cpu}.

\subsection{Particle dynamics on the CPU}\label{particle_dynamics}
The first step of the \gls{pd} simulation on the \gls{cpu} is the pre-force integration of the velocities to update the particle positions (\cref{pre_force_integration}). The latter particle movement requires synchronization between the \glspl{cpu} to account for the position update, which potentially moves particles from one block to another, making other \glspl{cpu} responsible for the particles. Computing particle-particle interactions by iterating over all particle pairs can quickly become very expensive due to its $O(n^2)$ complexity. Therefore, we insert the particles into linked cells such that iterating over the particle pairs is limited to neighboring linked cells. The linked cell size limits the maximum distance for which correct particle-particle interactions can be ensured (collision, lubrication correction). The linked cells have a size of 1.01 times the particle diameter, which is close to the smallest size that still leads to a correct collision detection.
Next, the lubrication correction routine is applied to all particle pairs with particles close to each other but not yet in contact (\cref{lubrication_correction}).
The particle-particle interactions are modeled using the \gls{dem} collision kernel (linear spring dashpot), which exerts forces and torques on overlapping particles.
The collision kernel needs history information from the previous time step, i.e., the accumulated tangential motion between the two colliding particles (\cref{history_information}). Since different processes may have handled the previous collisions, reducing the collision histories between the \glspl{cpu} is necessary, i.e., collecting all collision histories for a given particle in one process.
Then, the hydrodynamic forces and torques exerted by the fluid on the particles and the gravitational force are added to the total force.
Since different processes may have added forces and torques to a given particle, those contributions have to be collected in one process (another \gls{cpu}-\gls{cpu} communication).
Then, the post-integration is applied to update the particle velocities (\cref{post_force_integration}). Communication can be omitted here because the velocities are unused until the subsequent communication in the upcoming sub-cycle. 
Typically, $j$ particle sub-cycles are performed per time step since the \gls{dem} requires a finer resolution in time than the \gls{lbm} for an accurate contact representation~\citep[][]{rettingerEfficientFourwayCoupled2022}.
After completing $j$ sub-cycles, the next time step starts with the fluid dynamics on the \gls{gpu}.\newline
Running the fluid dynamics and coupling on the \gls{gpu} first and the particle dynamics on the \gls{cpu} second seems to be a promising candidate for overlapping the \gls{cpu} and \gls{gpu} computations to gain some performance improvements. Therefore, we elaborate on this possibility in the remainder of this section.
The following will refer to the different simulation modules as they are named in \cref{fig:implementation}.\newline
Under the condition that the numerical error must not increase, only some parts of the \gls{cpu} and \gls{gpu} computations can overlap, others cannot due to the dependencies of the two-way coupling. 
There are two ways of overlapping the \gls{cpu} and \gls{gpu} parts: within a time step or between subsequent time steps.
Only the post-force integration step of the particle simulation on the \gls{cpu} in time step $n$ depends on the reduction of the hydrodynamic forces on the \gls{gpu} in time step $n$, and therefore, the steps from pre-force integration to the application of external forces can, in principle, be overlap with the \gls{gpu} part.
The particle mapping on the \gls{gpu} at the beginning of time step $n+1$ requires the updated particle positions of the pre-force integration in time step $n$. Therefore, the steps from inserting the particles into linked cells until the post-force integration in time step $n$ can, in principle, overlap with the particle mapping on the \gls{gpu} in time step $n+1$.
However, the particle simulation typically consists of several sub-cycles (i.e., the \gls{cpu} part in \cref{fig:implementation} is executed several times per time step), and only the respective parts of the first and last sub-cycle can be overlapped. Therefore, the maximum achievable speedup due to overlapping depends on the number of sub-cycles. In accordance with the literature~\citep[][]{rettingerEfficientFourwayCoupled2022}, we use ten sub-cycles, which could potentially decrease the run time of the particle simulation on the \gls{cpu} by 20\% if the first and last sub-cycles could be entirely overlapped. However, this is an optimistic assumption because the interruption of the consecutive particle sub-cycles by the overlapping might negatively affect the possibility of caching particle data between consecutive sub-cycles.\newline
If one disregards the requirement that numerical error must not increase, one could use results from previous time steps, allowing both the fluid simulation on the \gls{gpu} and the particle simulation on the \gls{cpu} to run entirely in parallel. This might result in acceptable minor errors for systems with negligible particle movements, but this is not generally applicable and would require a detailed error analysis.

\section{Performance analysis}\label{performance}
We use the Juwels Booster cluster for the performance evaluation. Each GPGPU node consists of four Nvidia A100 40 GB \glspl{gpu}, two AMD EPYC 7402 \glspl{cpu} (24 cores per chip), and eight NUMA domains. Thus, each \gls{cpu} is shared by two \glspl{gpu} and is divided into four NUMA domains, with six cores belonging to one NUMA domain. Four out of the eight NUMA domains have independent PCIe lanes to the four \glspl{gpu} on Juwels Booster, ensuring that \gls{cpu}-\gls{gpu} communications do not interfere between the \glspl{gpu}. The following will refer to a \gls{gpu} and an associated (i.e., directly connected via a PCIe lane) NUMA domain with its six cores as a \gls{cpu}-\gls{gpu} pair. All \glspl{gpu} within a node are connected via NVLinks, allowing direct \gls{gpu}-\gls{gpu} communication. For communication between \glspl{gpu} not sharing a node, PCIe lanes connect each \gls{gpu} to its own Mellanox HDR200 InfiniBand ConnectX 6 adapter. A SCALE kernel (i.e., a 1:1 read/write ratio) yields a memory bandwidth of about 1400 GB/s for the A100 40 GB~\citep[][]{ernstAnalyticalPerformanceEstimation2023}. 
We use 20 cells per diameter to geometrically resolve the particles~\citep[][]{rettingerRheologyMobileSediment2022, rettingerEfficientFourwayCoupled2022, biegertCollisionModelGrainresolving2017, costaCollisionModelFully2015}.
The upcoming sections first introduce the computational properties of the simulated cases, followed by their performance results.

\subsection{Simulation setups}\label{setups}
\begin{figure}
  \centering
  \subfloat[\centering Dilute case]{\includegraphics[scale=0.5, height=10cm]{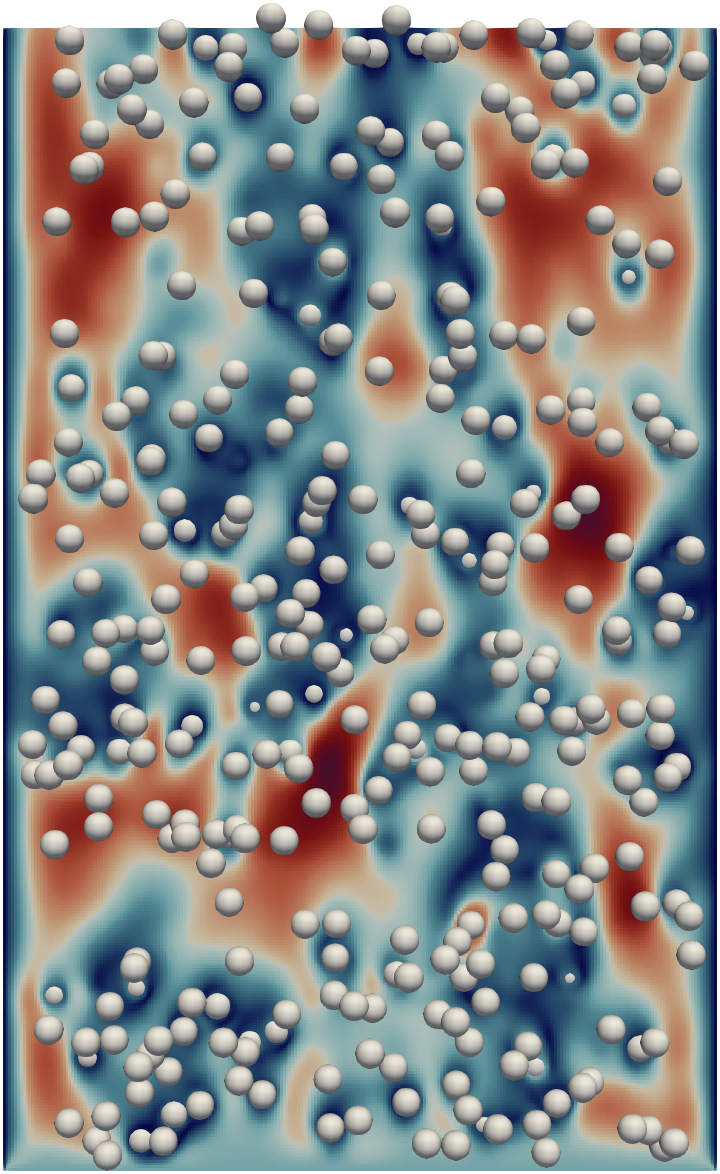}\label{fig:fluidized_bed_dilute}}
  \qquad
  \subfloat[\centering Dense case]{\includegraphics[scale=0.5, height=10cm]{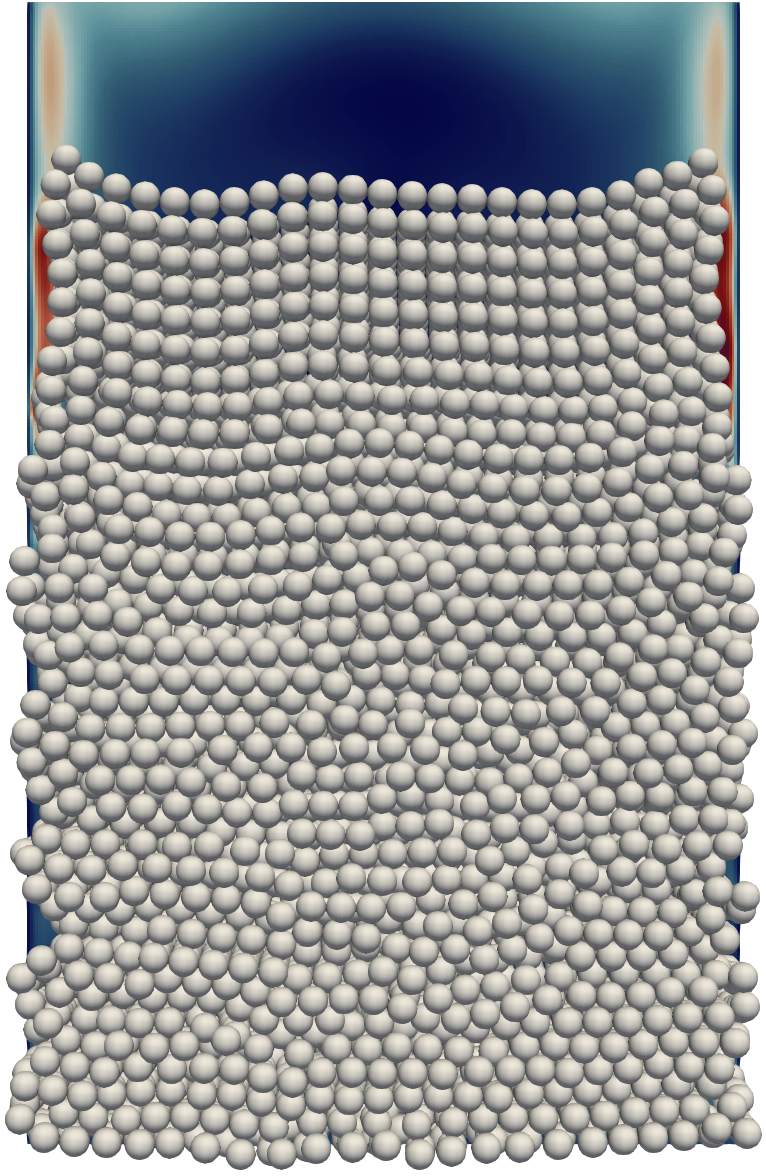}\label{fig:fluidized_bed_dense}}
  \caption{Visualization of the consolidated fluidized bed setup running on one \gls{cpu}-\gls{gpu} pair. For the fluid field, only a two-dimensional slice is visualized.}
\end{figure}

In order to study the performance of the here introduced hybrid coupled fluid-particle implementation, we are using a fluidized bed simulation. We compare two cases: the dilute case and the dense case. They exhibit different characteristics regarding the number of particles per volume and the number of particle-particle interactions. We choose these two cases to investigate how different particle workloads on the \glspl{cpu} influence the overall performance of the hybrid \gls{cpu}-\gls{gpu} implementation.
We use ten particle sub-cycles (\cref{implementation}) per time step.
We discretize the domain using $500 \times 200 \times 800 = 80 \times 10^6$ fluid cells. The velocity (inflow) \gls{bc} on the bottom and pressure (outflow) \gls{bc} on top of the domain govern the fluid dynamics. The remaining four boundaries are no-slip conditions ensuring a zero fluid velocity, i.e., the domain is not periodic. The particle Reynolds number is $1.0$, the Galileo number is around $8.9$, the gravitational acceleration is $9.81\ \text{m}/\text{s}^2$, and the particle fluid density ratio is $1.1$.
Planes surrounding the domain prevent the particles from leaving the domain by acting as walls that form a box.
The dilute case contains 627 particles. \cref{fig:fluidized_bed_dilute} illustrates the dilute setup.
Due to the low particle concentration, the effort for computing the particle-particle interactions (collisions and lubrication corrections) is low in the dilute case. 
The dense case is generally the same setup as the dilute case, except that the particle concentration is significantly higher limiting the fluidization (\cref{fig:fluidized_bed_dense}), resulting in 8073 particles, almost 13 times more than in the dilute case.

\subsection{Performance results}
In this section, we present and analyze the performance results. We first look at the individual run times of the different simulation modules to understand the bottlenecks.
Furthermore, we present a weak scaling benchmark for both cases up to 1024 \gls{cpu}-\gls{gpu} pairs.
\change{In addition, we show strong scaling results for three different problem sizes up to 1024 \gls{cpu}-\gls{gpu} pairs.}
Finally, we demonstrate the acceleration potential of hybrid implementations by comparing it to a large-scale \gls{cpu}-only simulation from the literature.
For all results, we average over 500 time steps.
In the following, we will refer to the performance criteria formulated in \cref{introduction}.

\subsubsection{Run times of different simulation modules}
\begin{figure*}
  \centering
  \pgfplotsset{/pgfplots/horizontal line legend/.style={legend image code/.code={\draw[ultra thick] (0cm,0cm)-- (.35cm,0cm);},},}

\begin{tikzpicture}
    \begin{axis} [ybar,
    ylabel=Time per time step (ms), ymin=0, ymax=45,
    symbolic x coords={PSM, comm, mapping, setU, redF, PD, other},
    xtick={PSM, comm, mapping, setU, redF, PD, other},
    xticklabel style={text height=2ex},
    legend style={at={(0.55,0.8)},anchor=east},
    legend cell align={left},
    grid, width=0.9*\textwidth, height=0.4*\textheight]
    \addplot [draw=matplotlibBlue,fill=matplotlibBlue!50, postaction={pattern=north east lines}, pattern color=black] coordinates {
        (PSM,24.78315357)
        (redF,4.912856478)
        (mapping,3.210133870259481)
        (setU,1.485454394)
        (PD,1.8600000000000003)
        (other,3.5557743957405172)
        (comm,0.10502729200000001)
    };
    \addplot [draw=matplotlibOrange,fill=matplotlibOrange!50, postaction={pattern=north east lines}, pattern color=black] coordinates {
        (PSM,31.901119082)
        (redF,18.391524202)
        (mapping,5.8220524171656685)
        (setU,2.792142046)
        (PD,41.96)
        (other,7.24992760483435)
        (comm,0.381234648)
        
    };
    \addplot[blue,horizontal line legend,sharp plot,update limits=false,ultra thick, dashed] coordinates { ([normalized]-0.4,23.9) ([normalized]0.4,23.9) };
    \addplot[black,horizontal line legend,sharp plot,update limits=false,ultra thick] coordinates { ([normalized]-0.4,17.4) ([normalized]0.4,17.4) };
    \legend{Dilute case, Dense case, PSM without particles, LBM roofline $T_\min$}
    \end{axis}
\end{tikzpicture}
  \caption{Individual run times of the different simulation modules on a \gls{cpu}-\gls{gpu} pair}\label{fig:runtimesa100}
\end{figure*}
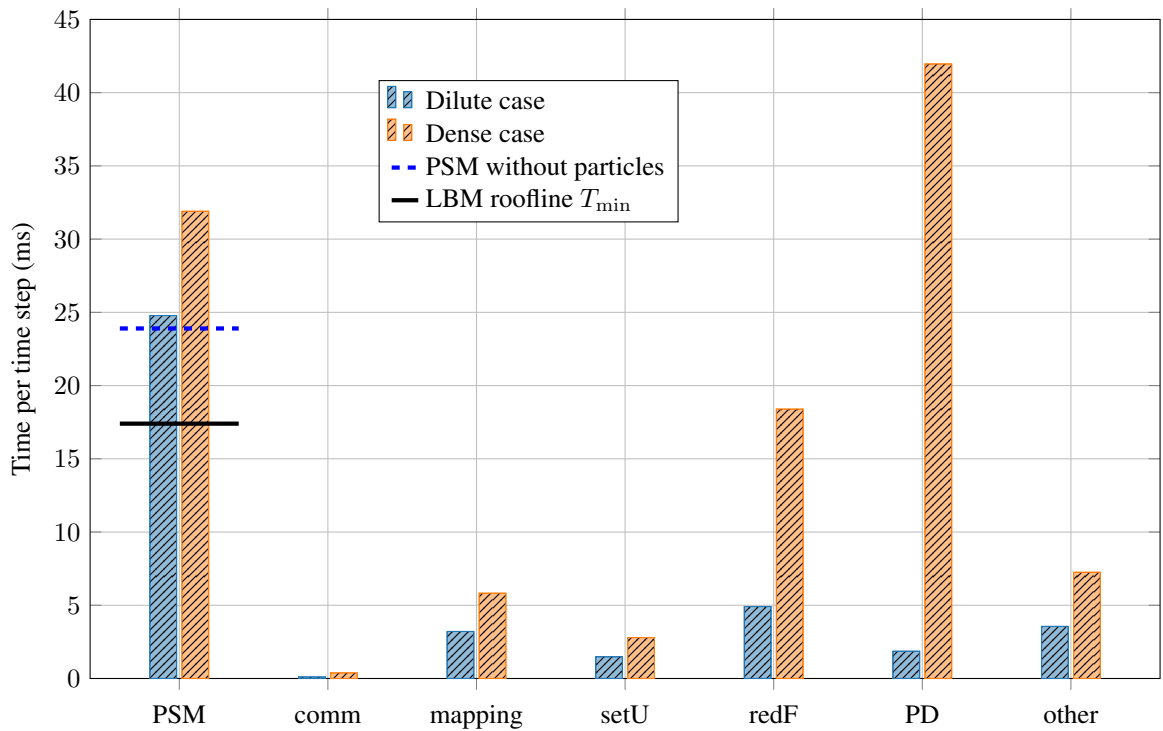

We investigate the run times of the different simulation modules to analyze the overhead introduced by the hybrid implementation, i.e., the \gls{cpu}-\gls{gpu} communication, assess the \gls{gpu} performance, and detect the overall bottlenecks. To evaluate the performance of the \gls{psm} kernel on the \gls{gpu}, we employ a roofline model~\citep[][]{hagerIntroductionHighPerformance2010} for the \gls{lbm} kernel.
We determine the maximal possible performance for the given kernel and hardware when exploiting the maximal memory bandwidth (i.e., the performance `lightspeed' estimation). The \gls{psm} kernel comprises the \gls{lbm} kernel plus additional memory transfers depending on the number of overlapping particles. As this number differs from cell to cell, the \gls{psm} roofline model would not be straightforward. Therefore, we analyze the \gls{lbm} model only, keeping in mind that this is a too optimistic performance estimation.
Since we use the D3Q19 lattice model, we read and write 19 \gls{pdfs} per cell and time step.
This results in 19 reads and 19 writes (double-precision), i.e., 304 bytes to update one lattice cell~\citep[][]{feichtingerPerformanceModelingAnalysis2015}. The domain consists of 8e7 fluid cells. This results in the following minimal run time per time step according to the roofline model:
\begin{equation}
  \begin{aligned}
    T_\min &= \frac{304 \ \text{B/cell} \cdot \text{8e7} \ \text{cells/time step}}{1400 \ \text{GB/s}} \\
           &= 17.4 \ \text{ms/time step}.
  \end{aligned}
\end{equation}
We divide the total run time into the following modules: the \gls{psm} kernel (PSM), the \gls{cpu}-\gls{gpu} communication (comm), the particle mapping (mapping), setting the particle velocities (setU), reducing the hydrodynamic forces $\boldsymbol{F}_{\text{p},i}^{\text{fp}}$ and torques $\boldsymbol{T}_{\text{p},i}^{\text{fp}}$ on the particles (redF), computing the \gls{pd} and finally the remaining tasks (other), e.g., the \gls{lbm} boundary handling.\ \cref{fig:runtimesa100} reports the run times per time step for these modules using a single \gls{cpu}-\gls{gpu} pair. Furthermore, a dashed horizontal line indicates the run time of the \gls{psm} kernel for a simulation without any particles in the domain. While this physically behaves precisely like an \gls{lbm} kernel, it allows the quantification of the overhead due to the different code structure of the \gls{psm} compared to the \gls{lbm} kernel without the additional effort due to particles inside the domain.
In the dilute case, the \gls{psm} kernel needs about 42\% more time per time step than the \gls{lbm} lightspeed estimation, and the dense case 83\% more time. Compared to the \gls{psm} kernel without particles, the dilute case needs about 4\% more time per time step, and the dense case 34\% more time. The \gls{cpu}-\gls{gpu} communication is negligible for both cases. All modules take longer in the dense case than in the dilute case. While in the dilute case, the \gls{psm} kernel accounts for the majority of the run time, the \gls{pd} simulation needs more time than the \gls{psm} kernel in the dense case. Still, most of the run time is spent on \gls{gpu} routines in the dense case.\newline
The overhead introduced by the hybrid implementation is negligible because we only transfer a small amount of double-precision values per particle but no fluid cells, and thanks to the problem-aware co-partitioning. The overhead, therefore, shows that a hybrid parallelization with the presented technique is a viable approach, and the first criterion is met. The performance of the \gls{psm} kernel is close to utilizing the total memory bandwidth of the A100, especially considering that the given roofline model does not take the memory traffic due to the solid part of the \gls{psm} kernel into account. Therefore, the second criterion is met. There is a significant performance gain for the \gls{psm} kernel on the \gls{gpu} compared to a \gls{cpu}-only implementation since the \gls{psm} kernel is utilizing almost the entire memory bandwidth, which is significantly lower on \glspl{cpu}.
Even in the dense case, the \gls{gpu} run time accounts for most of the total time because the fluid and coupling workload is much higher than the particle workload, even though we use ten particle sub-cycles. Therefore, the third criterion is met.
The \gls{psm} kernel exhibits two remarkable effects that we analyze in detail below using NVIDIA Nsight Compute\footnote{\url{https://developer.nvidia.com/nsight-compute}}.\newline
First, the significant difference between the \gls{lbm} main memory roofline (horizontal line) and the \gls{psm} kernel without particles (dashed horizontal line) in \cref{fig:runtimesa100} indicates that the \gls{psm} kernel cannot fully utilize the main memory bandwidth as the memory transfers of the \gls{psm} kernel in the absence of particles are very similar to the ones assumed in the \gls{lbm} main memory roofline. This is noticeable because the \gls{psm} is an extension of the \gls{lbm}, and it is known that the \gls{lbm} can nearly fully utilize the main memory bandwidth on the A100 architecture~\citep[][]{lehmannAccuracyPerformanceLattice2022, holzerDevelopmentCentralmomentPhasefield2024a}.
The \gls{psm} kernel uses 196 registers per thread. This large number of registers per thread heavily limits the number of warps per \gls{sm}, leading to a maximum possible occupancy of the \gls{sm}s of only 12.50\%. This low occupancy is insufficient to issue enough load/store instructions to fully exploit the main memory bandwidth, resulting in a difference between the \gls{psm} without particles and the \gls{lbm} memory roofline.\newline
\change{Second, while the \gls{psm} kernel is in the dilute case only slightly slower than without particles, this difference gets more significant in the dense case.} Several effects differentiate the dilute and dense cases.
First, more particles increase the unfavorable \gls{gpu} workload, i.e., warp divergence and coalesced access. The branch instructions increase by 2.06\% in the dilute case and by 18.18\% in the dense case compared to the \gls{psm} kernel without particles. The number of excessive sectors due to uncoalesced global accesses increases by 78.49\% from the dilute case to the dense case resulting in 21\% of the total sectors in the dense case.
Besides that, the number of instructions issued increases by 4.32\% in the dilute case and 39.51\% in the dense case, which is in a similar range than the run time increase, corresponding to the instruction boundness of the code mentioned in the first effect (i.e., the low occupancy of the \gls{sm}s limits the issued instructions).
Since the increase in main memory traffic is 0.98\% (dilute case) and 12.15\% (dense case) and therefore smaller than the run time increase, the main memory utilization drops from 65.85\% (dilute case) to 52.56\% (dense case).

\subsubsection{Weak scaling}\label{weak_scaling}
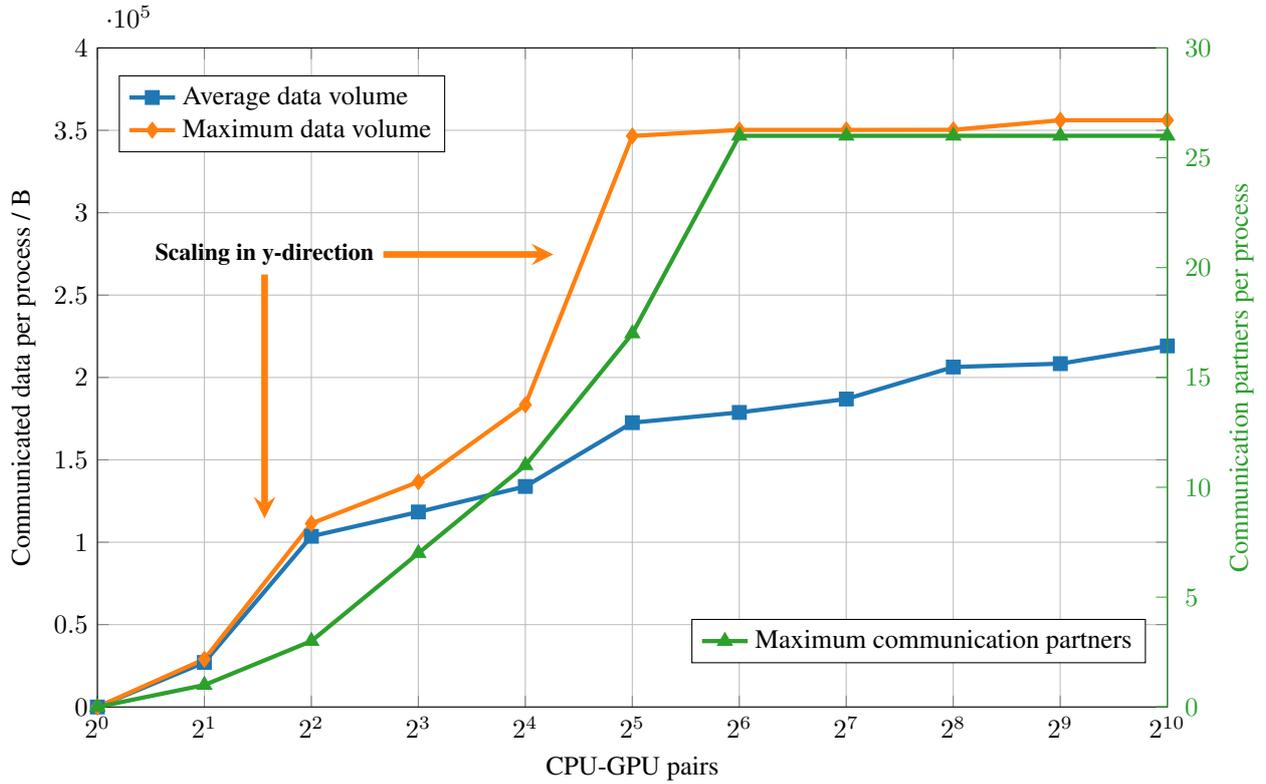
\begin{figure*}
  \centering
  \begin{tikzpicture}
  \begin{semilogxaxis}[
      xlabel=CPU-GPU pairs,
      ylabel=Communicated data per process / B,
      log basis x={2},
      xmin=1,
      xmax=1024,
      ymin=0,
      ymax=4e5,
      legend style={at={(0.02,0.9)},anchor=west},
      legend cell align={left},
      width=0.9*\textwidth,
      height=0.4*\textheight,
      grid
      ]

  \addplot[mark=square*, mark size=2pt, ultra thick, matplotlibBlue] plot coordinates {
    (1,0.0)
    (2,27033.14481818182)
    (4,103601.08259090909)
    (8,118363.93193181818)
    (16,133842.30953409092)
    (32,172566.88514772727)
    (64,178778.78458806817)
    (128,186917.6586747159)
    (256,206344.17667116478)
    (512,208371.74487109375)
    (1024,219141.2755271662)
  };
  \addlegendentry{Average data volume}

  \addplot[mark=diamond*, mark size=2pt, ultra thick, matplotlibOrange] plot coordinates {
    (1,0)
    (2,28998.76109090909)
    (4,111370.6989090909)
    (8,136642.25145454545)
    (16,183419.76763636363)
    (32,346582.964)
    (64,350245.50872727274)
    (128,350245.444)
    (256,350318.08327272726)
    (512,356112.12963636365)
    (1024,356112.12963636365)
  };
  \addlegendentry{Maximum data volume}
  \end{semilogxaxis}

  \begin{semilogxaxis}[
    axis y line*=right,
    axis x line=none,
    y axis line style = {matplotlibGreen},
    ytick style = {matplotlibGreen}, 
    ylabel style = {matplotlibGreen},
    yticklabel style = {matplotlibGreen, xshift=3pt},
    xlabel=CPU-GPU pairs,
    ylabel=Communication partners per process,
    log basis x={2},
    xmin=1,
    xmax=1024,
    ymin=0,
    ymax=30,
    legend style={at={(0.98,0.1)},anchor=east},
    legend cell align={left},
    width=0.9*\textwidth,
    height=0.4*\textheight,
    ]

  \addplot[mark=triangle*, mark size=2pt, ultra thick, matplotlibGreen] plot coordinates {
    (1,0)
    (2,1.0)
    (4,3.0)
    (8,7.0)
    (16,11.0)
    (32,17.0)
    (64,26.0)
    (128,26.0)
    (256,26.0)
    (512,26.0)
    (1024,26.0)
  };
  \addlegendentry{Maximum communication partners}
\end{semilogxaxis}
\node[font=\small] (text) at (2.2,6) {\textbf{Scaling in y-direction}};
\draw[->, line width=2.5pt, >=stealth, matplotlibOrange] (text) -- (6,6);
\draw[->, line width=2.5pt, >=stealth, matplotlibOrange] (text) -- (2.2,2.5);
\end{tikzpicture}
  \caption{Weak scaling of the particle synchronization in the dense case up to 1024 \gls{cpu}-\gls{gpu} pairs. Both the maximum and average amount of data (in bytes) a process either sends to or receives from its neighboring processes per synchronization call are illustrated (left y-axis), and the maximum number of synchronization partners (right y-axis).}\label{fig:weak_scaling_communication_dense}
\end{figure*}

When increasing the simulation domain further to simulate physically relevant scenarios, using a single \gls{cpu}-\gls{gpu} pair is often insufficient. Instead, multiple pairs or even multiple nodes of a supercomputer must be used. Therefore, a satisfactory weak scaling is desirable. For weak scaling, the problem size is increased with an increasing number of \gls{cpu}-\gls{gpu} pairs, keeping the workload per \gls{cpu}-\gls{gpu} pair constant. When having a perfect weak scaling, the performance per \gls{cpu}-\gls{gpu} pair stays constant, independent of the number of \gls{cpu}-\gls{gpu} pairs used. In the context of \gls{lbm}, MLUPs is a standard performance metric for weak scaling~\citep[][]{holzerHighlyEfficientLattice2021}, meaning how many lattice cell updates the hardware performs per second. 
We use the total run time for computing the MLUPs, containing both the \gls{cpu} (particles) and the \gls{gpu} (fluid, coupling) time. For the weak-scaling plots, we have conducted at least three benchmarking runs and will use the best sample in the following.
We start with a single \gls{cpu}-\gls{gpu} pair and a single domain block as described in \cref{setups}. We then double the number of \gls{cpu}-\gls{gpu} pairs succesively until we reach 1024. At the same time, we double the domain blocks/size alternately in each direction: $2\times 1 \times 1$ blocks (two \glspl{gpu}), $2\times 2 \times 1$ blocks (four \glspl{gpu}), $2\times 2 \times 2$ (eight \glspl{gpu}), etc.\ \cref{fig:weak_scaling} and \cref{fig:weak_scaling_efficiency} report the weak scaling for both cases. \change{Dashed lines indicate the ideal curves.} To our best knowledge, this is the most extensive weak scaling of a hybrid fluid-particle implementation presented in the literature.
\begin{figure*}
  \centering
  \begin{tikzpicture}
    \begin{semilogxaxis}[
        xlabel=CPU-GPU pairs,
        ylabel=MLUPs per pair,
        log basis x={2},
        xmin=1,
        xmax=1024,
        ymin=0,
        ymax=2200,
        legend style={at={(0.03,0.5)},anchor=west},
        legend cell align={left},
        width=0.9*\textwidth,
        height=0.4*\textheight,
        grid
        ]
    \addplot[mark=*, mark size=2pt, ultra thick, matplotlibBlue] plot coordinates {
      (1,2027.3)
      (2,1963.23)
      (4,1814.74)
      (8,1756.03)
      (16,1712.99)
      (32,1599.67)
      (64,1537.87)
      (128,1525.81)
      (256,1493.45)
      (512,1481.59)
      (1024,1435.99)
    };
    \addplot[mark=diamond*, mark size=2pt, ultra thick, matplotlibOrange] plot coordinates {
      (1,706.613)
      (2,634.843)
      (4,568.501)
      (8,549.758)
      (16,519.1)
      (32,470.161)
      (64,476.227)
      (128,436.286)
      (256,377.213)
      (512,414.443)
      (1024,377.581)
    };
    \addplot[blue, ultra thick, dashed] coordinates { (1,2027.3) (1024,2027.3) };
    \addplot[blue, ultra thick, dashed] coordinates { (1,706.613) (1024,706.613) };
    \addlegendentry{Dilute case}
    \addlegendentry{Dense case}
    \addlegendentry{Ideal}
    \end{semilogxaxis}
\end{tikzpicture}
  \caption{Weak scaling performance for both cases up to 1024 \gls{cpu}-\gls{gpu} pairs}\label{fig:weak_scaling}
  \vspace{0.5cm}
\end{figure*}
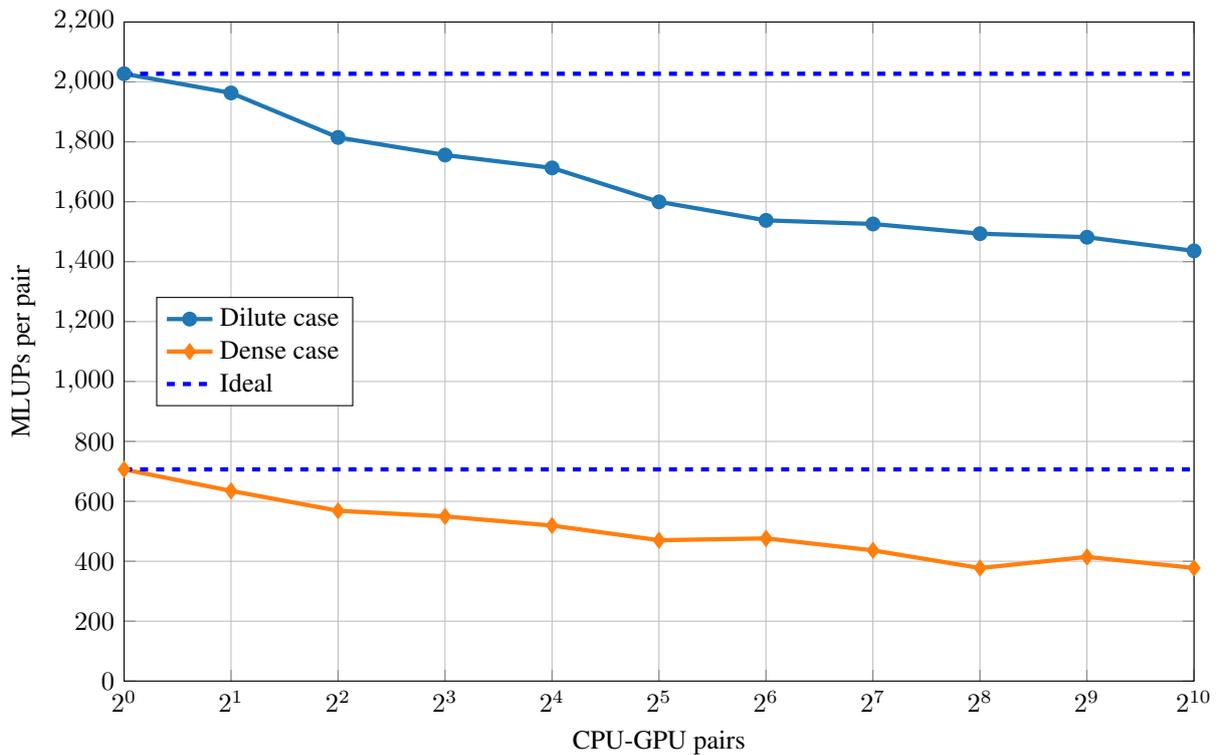
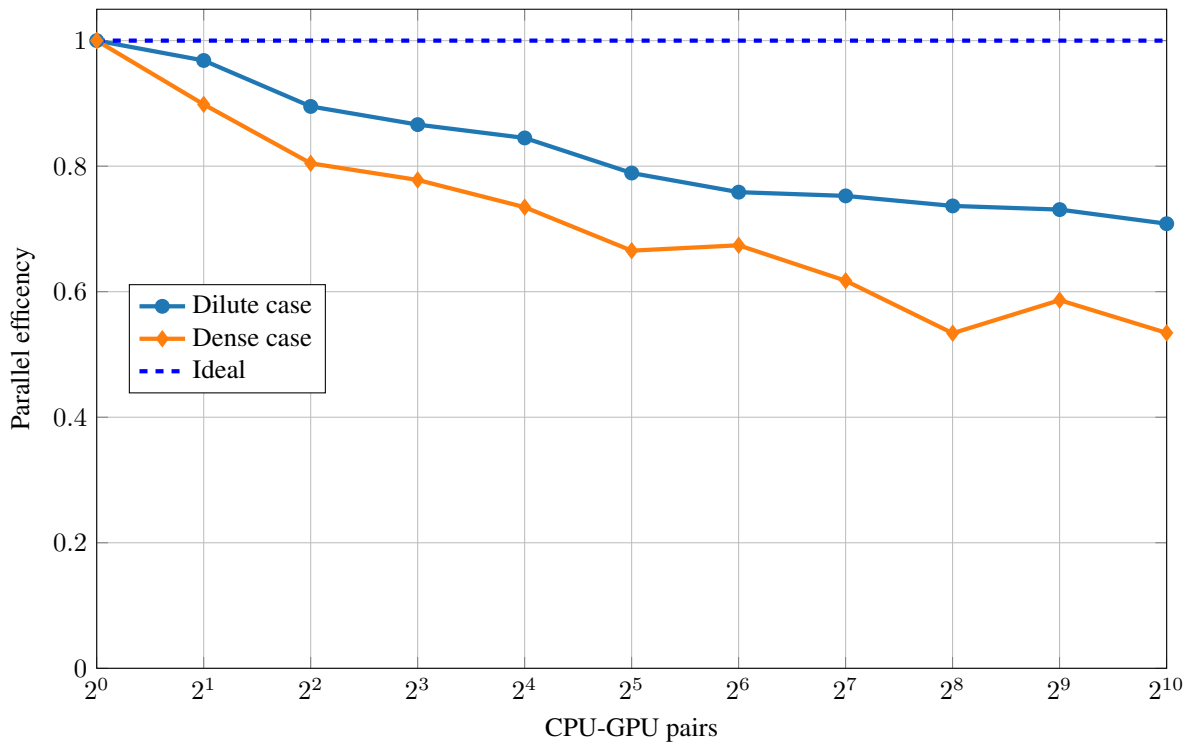
\begin{figure*}
  \centering
  \begin{tikzpicture}
    \begin{semilogxaxis}[
        xlabel=CPU-GPU pairs,
        ylabel=Parallel efficency,
        log basis x={2},
        xmin=1,
        xmax=1024,
        ymin=0,
        ymax=1.05,
        legend style={at={(0.03,0.5)},anchor=west},
        legend cell align={left},
        width=0.9*\textwidth,
        height=0.4*\textheight,
        grid
        ]
    \addplot[mark=*, mark size=2pt, ultra thick, matplotlibBlue] plot coordinates {
      (1,2027.3/2027.3)
      (2,1963.23/2027.3)
      (4,1814.74/2027.3)
      (8,1756.03/2027.3)
      (16,1712.99/2027.3)
      (32,1599.67/2027.3)
      (64,1537.87/2027.3)
      (128,1525.81/2027.3)
      (256,1493.45/2027.3)
      (512,1481.59/2027.3)
      (1024,1435.99/2027.3)
    };
    \addplot[mark=diamond*, mark size=2pt, ultra thick, matplotlibOrange] plot coordinates {
      (1,706.613/706.613)
      (2,634.843/706.613)
      (4,568.501/706.613)
      (8,549.758/706.613)
      (16,519.1/706.613)
      (32,470.161/706.613)
      (64,476.227/706.613)
      (128,436.286/706.613)
      (256,377.213/706.613)
      (512,414.443/706.613)
      (1024,377.581/706.613)
    };
    \addplot[blue, ultra thick, dashed] coordinates { (1,1) (1024,1) };
    \addlegendentry{Dilute case}
    \addlegendentry{Dense case}
    \addlegendentry{Ideal}
    \end{semilogxaxis}
\end{tikzpicture}
  \caption{Weak scaling parallel efficiency for both cases up to 1024 \gls{cpu}-\gls{gpu} pairs}\label{fig:weak_scaling_efficiency}
\end{figure*}
We observe a roughly three times higher performance for the dilute case than for the dense case. Both cases show a parallel efficiency decrease particularly strong in the beginning. The parallel efficiency is 71\% in the dilute case and 53\% in the dense case when using 1024 \gls{cpu}-\gls{gpu} pairs, which corresponds to a domain size of $8000 \times 1600 \times 6400 = 8.192 \times 10^{10}$ fluid cells. A similar scaling behavior has been observed in the literature both for other hybrid~\citep[][]{ kotsalosDigitalBloodMassively2021} and \gls{cpu}-only fluid-particle implementations~\citep[][]{rettingerFullyResolvedSimulations2017}.\newline
Interpreting the overall weak scaling behavior requires more detailed analysis of the scaling of the different simulation modules. When using a single \gls{cpu}-\gls{gpu} pair, the dominating routines are the \gls{pd}, the \gls{psm} kernel, and the coupling (i.e., particle mapping, setting the particle velocities and reducing the hydrodynamic forces $\boldsymbol{F}_{\text{p},i}^{\text{fp}}$ and torques $\boldsymbol{T}_{\text{p},i}^{\text{fp}}$
on the particles). Additionally, we must now consider the communication (comm) overhead that arises from using multiple \gls{cpu}-\gls{gpu} pairs. On the one hand, this is the \gls{pd} communication (\gls{cpu}-\gls{cpu} communication), on the other hand, the PSM communication (\gls{gpu}-\gls{gpu} communication). \cref{fig:weak_scaling_routines_dilute} and \cref{fig:weak_scaling_routines_dense} show the weak scaling behavior of the dominating simulation modules for both cases. The communication numbers cover the communication steps themselves, but also load imbalances between two communication steps.
\begin{figure*}
  \centering
  \begin{tikzpicture}[scale=0.9]
  \begin{semilogxaxis}[
      xlabel=CPU-GPU pairs,
      ylabel=Time per time step per pair (ms),
      log basis x={2},
      xmin=1,
      xmax=1024,
      ymin=0,
      ymax=30,
      legend style={at={(1.03,0.5)},anchor=west},
      legend cell align={left},
      width=0.9*\textwidth,
      height=0.4*\textheight,
      grid
      ]

  \addplot[mark=square*, mark size=2pt, ultra thick, matplotlibRed] plot coordinates {
    (1,24.58)
    (2,24.75)
    (4,25.43)
    (8,25.4475)
    (16,25.463749999999997)
    (32,25.51875)
    (64,25.534375)
    (128,25.601875)
    (256,25.623984375000003)
    (512,25.643632812499998)
    (1024,25.65595703125)
  };
  \addlegendentry{PSM}

  \addplot[mark=square, mark size=2pt, ultra thick, matplotlibBrown] plot coordinates {
    (1,0.0)
    (2,0.24)
    (4,0.515)
    (8,0.595)
    (16,1.13875)
    (32,1.8675)
    (64,1.9853125)
    (128,2.2434375)
    (256,2.273828125)
    (512,2.34953125)
    (1024,2.50908203125)
  };
  \addlegendentry{PSM comm}

  \addplot[mark=oplus*, mark size=2pt, ultra thick, matplotlibGreen] plot coordinates {
    (1,1.84)
    (2,1.83)
    (4,2.6599999999999997)
    (8,2.735)
    (16,2.79875)
    (32,3.14)
    (64,3.128125)
    (128,3.20453125)
    (256,3.324609375)
    (512,3.3340234375)
    (1024,3.4020898437500002)
  };
  \addlegendentry{PD}

  \addplot[mark=o, mark size=2pt, ultra thick, matplotlibPurple] plot coordinates {
    (1,0.02)
    (2,0.44)
    (4,1.71)
    (8,2.3125)
    (16,2.9425)
    (32,5.456875)
    (64,7.079375)
    (128,6.977656250000001)
    (256,8.04328125)
    (512,8.4655078125)
    (1024,9.85158203125)
  };
  \addlegendentry{PD comm}

  \addplot[mark=triangle*, mark size=2pt, ultra thick, matplotlibPink] plot coordinates {
    (1,10.54)
    (2,11.129999999999999)
    (4,11.89)
    (8,12.68)
    (16,12.80625)
    (32,12.762500000000001)
    (64,12.8503125)
    (128,12.837968750000002)
    (256,12.896171875)
    (512,12.98875)
    (1024,13.045175781249998)
  };
  \addlegendentry{Coupling}

  \end{semilogxaxis}
\end{tikzpicture}
  \caption{Weak scaling performance of the dominating modules for the dilute case up to 1024 \gls{cpu}-\gls{gpu} pairs}\label{fig:weak_scaling_routines_dilute}
  \vspace{0.5cm}
\end{figure*}
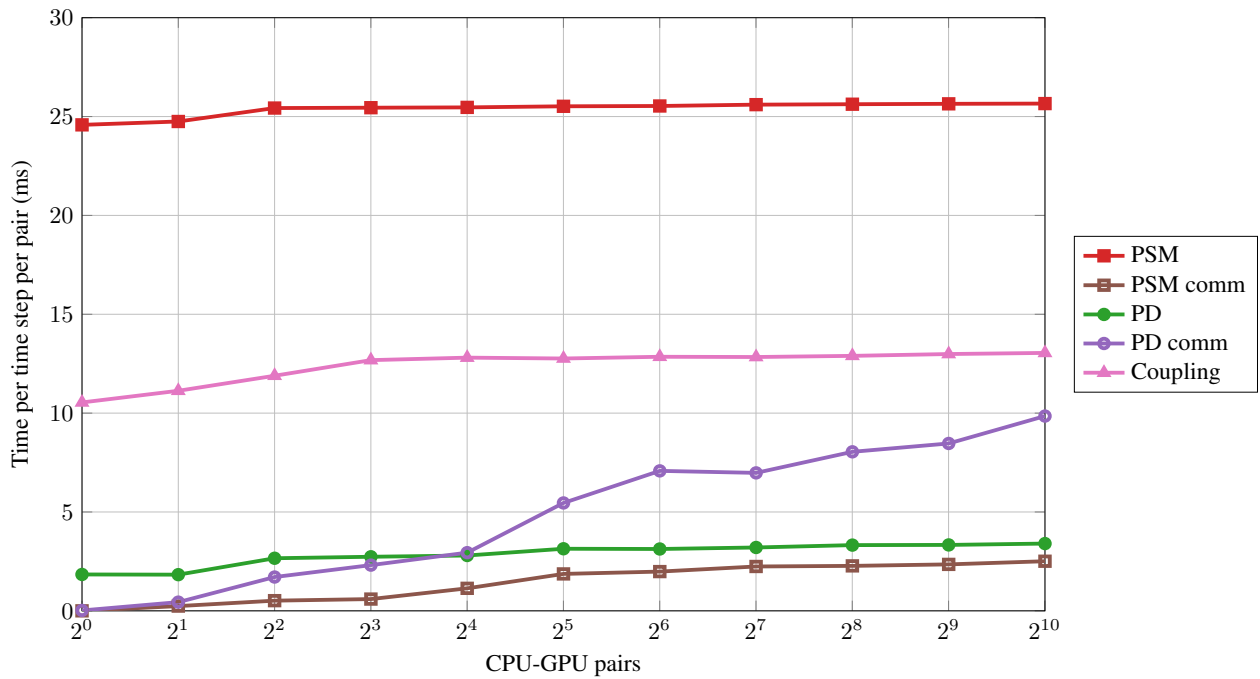
\begin{figure*}
  \centering
  \begin{tikzpicture}[scale=0.9]
  \begin{semilogxaxis}[
      xlabel=CPU-GPU pairs,
      ylabel=Time per time step per pair (ms),
      log basis x={2},
      xmin=1,
      xmax=1024,
      ymin=0,
      ymax=70,
      legend style={at={(1.03,0.5)},anchor=west},
      legend cell align={left},
      width=0.9*\textwidth,
      height=0.4*\textheight,
      grid
      ]

  \addplot[mark=square*, mark size=2pt, ultra thick, matplotlibRed] plot coordinates {
    (1,31.64)
    (2,33.019999999999996)
    (4,33.63)
    (8,33.865)
    (16,34.16375)
    (32,34.491875)
    (64,34.560625)
    (128,34.55421875)
    (256,38.04453125)
    (512,35.637578125000005)
    (1024,38.15447265625)
  };
  \addlegendentry{PSM}

  \addplot[mark=square, mark size=2pt, ultra thick, matplotlibBrown] plot coordinates {
    (1,0.0)
    (2,0.46)
    (4,0.645)
    (8,0.6975)
    (16,1.2125)
    (32,2.066875)
    (64,2.29125)
    (128,2.4053125)
    (256,3.113984375)
    (512,2.821328125)
    (1024,3.0503515625)
  };
  \addlegendentry{PSM comm}

  \addplot[mark=oplus*, mark size=2pt, ultra thick, matplotlibGreen] plot coordinates {
    (1,45.459999999999994)
    (2,50.09)
    (4,56.095)
    (8,55.8)
    (16,56.74125000000001)
    (32,59.916250000000005)
    (64,59.494687500000005)
    (128,60.9278125)
    (256,62.697578125)
    (512,61.9122265625)
    (1024,62.57080078125001)
  };
  \addlegendentry{PD}

  \addplot[mark=o, mark size=2pt, ultra thick, matplotlibPurple] plot coordinates {
    (1,1.96)
    (2,6.42)
    (4,11.715)
    (8,14.56)
    (16,20.802500000000002)
    (32,28.955)
    (64,27.888437500000002)
    (128,42.043125)
    (256,58.32515625)
    (512,46.796992187499995)
    (1024,58.0169140625)
  };
  \addlegendentry{PD comm}

  \addplot[mark=triangle*, mark size=2pt, ultra thick, matplotlibPink] plot coordinates {
    (1,31.8)
    (2,33.580000000000005)
    (4,36.39)
    (8,38.315)
    (16,38.7025)
    (32,40.681875000000005)
    (64,40.11875)
    (128,40.223125)
    (256,46.436562499999994)
    (512,42.322070312499996)
    (1024,46.645312499999996)
  };
  \addlegendentry{Coupling}

  \end{semilogxaxis}
\end{tikzpicture}
  \caption{Weak scaling performance of the dominating modules for the dense case up to 1024 \gls{cpu}-\gls{gpu} pairs}\label{fig:weak_scaling_routines_dense}
\end{figure*}
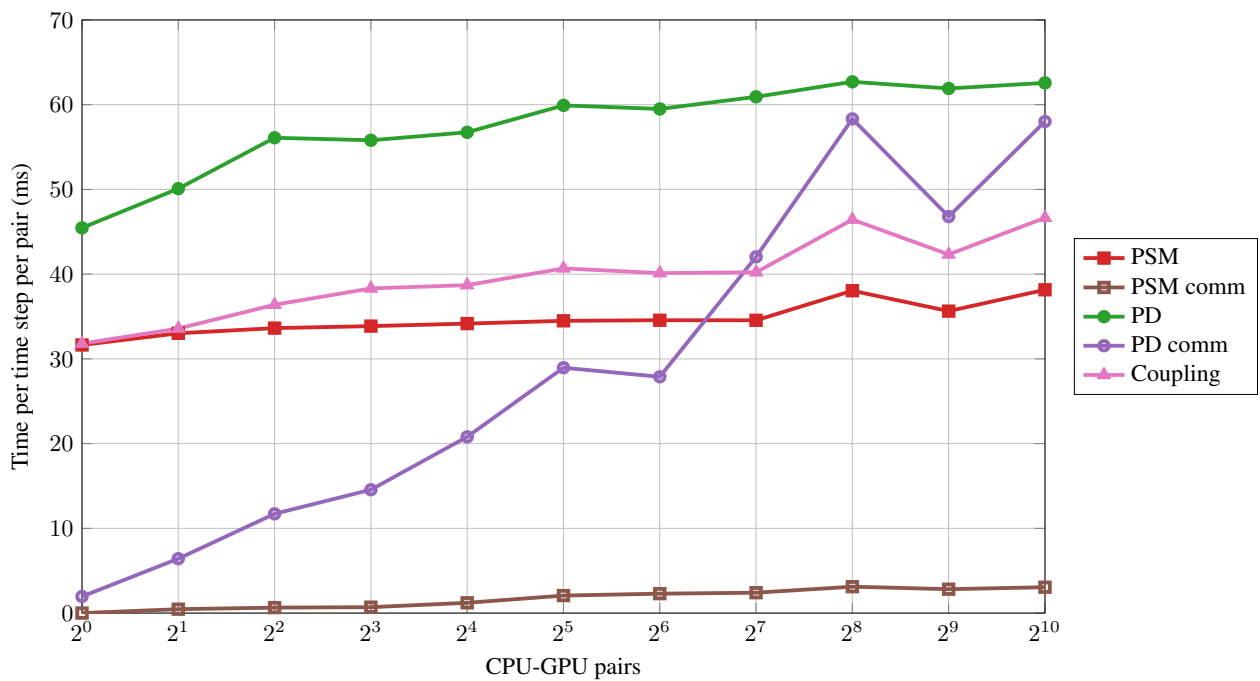
The different modules show similar qualitative scaling behavior when comparing the two cases. The \gls{psm} kernel scales quite well in both cases. The corresponding \gls{gpu}-\gls{gpu} communication (\gls{psm} comm) is negligible. The \gls{pd} run time increases initially and then shows saturation. The \gls{cpu}-\gls{cpu} communication (\gls{pd} comm) increases drastically, surpassing the run time of the \gls{psm} kernel and the coupling in the dense case. The \gls{pd} and the corresponding communication are more relevant for the overall scaling in the dense case than in the dilute case. The coupling scales similarly to the \gls{pd}.\newline
The \gls{psm} workload per \gls{gpu} stays constant in the weak scaling explaining the nearly perfect scaling.
Since we are hiding the \gls{psm} communication (\cref{implementation}), we expect it to be negligible.
We expect the \gls{pd} and the coupling run time to increase initially because the number of neighboring blocks increases. More neighboring blocks lead to more ghost particles per block, resulting in a higher workload. This effect decreases when blocks have neighbors in all directions, resulting in an almost linear scaling from this point on. This phenomenon has been reported in the literature~\citep[][]{rettingerFullyResolvedSimulations2017}.
The methodology requires ten particle sub-cycles per time step and three communication steps per sub-cycle for a physically accurate simulation. Additionally, the simulation requires two \gls{cpu}-\gls{cpu} communication steps per time step apart from the sub-cycles (\cref{fig:implementation}). We have 32 \gls{cpu}-\gls{cpu} communication steps per time step, which cannot be hidden behind other routines. This becomes the dominating factor for the decrease of the overall weak scaling performance in both cases.\newline
To gain more profound insights into the reasons for this significant increase of the \gls{pd} comm time in the weak scaling and to estimate the scaling behavior beyond 1024 \gls{cpu}-\gls{gpu} pairs, we investigate the data transfers of the \gls{pd} comm in more detail for the dense case. The \gls{pd} comm consists of two parts: the synchronization of particle quantities between processes and the reduction of particle quantities (\cref{particle_dynamics}). In the following, we focus on the dominating part, the synchronization. \cref{fig:weak_scaling_communication_dense} plots the maximum and average amount of data a process either sends to or receives from its neighboring processes per synchronization call. The two arrows indicate the steps in which the weak scaling doubles the domain in the y-direction (the out-of-plane direction normal to the cross-section in \cref{fig:fluidized_bed_dilute}) for the first two times. Furthermore, the figure illustrates the maximum number of synchronization (i.e., communication) partners per process.
The maximum number of communication partners increases initially, reaching 26 starting from $2^6$ processes and then staying constant.
The maximum amount of data a process sends to or receives from its neighboring processes increases to $2^5$ processes and then saturates.
The maximum data volume per process already saturates one process earlier than the maximum number of communication partners.
The average data volume per process increases until $2^{10}$ processes. However, the increase from $2^0$ to $2^5$ processes is larger than from $2^5$ to $2^{10}$.
For both the maximum and average data curves, the most significant spikes can be observed when going from $2^1$ to $2^2$ and from $2^4$ to $2^5$ processes, i.e., when the weak scaling doubles the domain size in the y-direction for the first two times.
The maximum and average communication data show correlation with each other, as well as with the \gls{pd} comm time in \cref{fig:weak_scaling_routines_dense}.\newline
The maximum number of communication partners is bounded from above at 26 because this is the maximum number of neighboring blocks in 3D (including blocks that only share a corner) because a $3\times 3 \times 3$ setup of blocks consists of the center block and 26 neighboring blocks. The weak scaling obtains this situation for the first time for $2^6$ processes since this corresponds to $4\times 4\times 4$ blocks, and therefore, at least one block is entirely surrounded by neighboring blocks.
The initial increase in maximum data transfer in the weak scaling is expected since the maximum number of neighbors increases. Therefore, more data has to be communicated per synchronization. The large spikes correspond to the cases where the domain is increased in the y-direction, which causes the most significant increase of the boundary layer between two blocks and, therefore, the most significant communication data increase. The step from $2^4$ to $2^5$ increases the maximum number of communication partners stronger than the step from $2^1$ to $2^2$. Therefore, the maximum data volume increases stronger accordingly (see the two arrows in \cref{fig:weak_scaling_communication_dense}).
As expected for nearest-neighbor communication, the maximum communication saturates if other processes entirely surround at least one process.
However, the maximum data volume per process already saturates one process earlier than the maximum number of communication partners is reached. This is because from $2^5$ to $2^6$ processes, the domain is increased in z-direction for the second time. However, due to the arrangement of the particles, synchronization only happens in positive z-direction. Therefore, doubling the domain in z-direction requires no additional communication since still, no process has to communicate in both z-directions.
The average data size still increases after the saturation of the maximum data because the ratio of boundary blocks (with less than 26 communication partners) compared to the total number of blocks decreases, i.e., proportionally, more blocks are becoming blocks with the maximum data size.
The maximum communication serves as an upper bound for the convergence of the average communication data.
Due to the pure nearest neighbor communication and the resulting convergence of the average data size per process, we expect promising scalability on larger systems beyond 1024 \gls{cpu}-\gls{gpu} pairs.
However, the scaling of the \gls{pd} comm depends strongly on the \gls{pd} methodology, its implementation, the available hardware, and the assignment of the blocks to the hardware.
Reducing the collision history is part of \gls{pd} comm, which includes swapping old and new contact information. Since this swap is also necessary without using multiple \gls{cpu}-\gls{gpu} pairs, \gls{pd} comm is bigger zero even when using a single \gls{cpu}-\gls{gpu} pair.
Overall, we observe a weak scaling performance that justifies using multiple supercomputer nodes. Therefore, the fourth criterion is met.

\subsubsection{Strong scaling}
\change{
For strong scaling, the problem size is fixed with an increasing number of \gls{cpu}-\gls{gpu} pairs, decreasing the workload per \gls{cpu}-\gls{gpu} pair. When having a perfect strong scaling, the run time decreases to the same extent as the number of \gls{cpu}-\gls{gpu} pairs increases.
We present strong scaling results for three problem sizes: small, medium, and large.
The small problem consists of $500 \times 200 \times 800$ fluid cells, as described in \cref{setups}, which fits on a single \gls{cpu}-\gls{gpu} pair. For the medium problem, the small problem size is doubled in all three directions, resulting in $1000 \times 400 \times 1600$ fluid cells, which fits on eight \gls{cpu}-\gls{gpu} pairs. For the large problem, the medium problem size is again doubled in all three directions, resulting in $2000 \times 800 \times 3200$ fluid cells, which fits on 64 \gls{cpu}-\gls{gpu} pairs. For each problem size, we successively double the number of \gls{cpu}-\gls{gpu} pairs four times in x-, y-, z-, and again in x-direction while keeping the problem size constant.
\cref{fig:strong_scaling} and \cref{fig:strong_scaling_efficiency} illustrate the strong scaling results employing both the dilute and dense case for the three different problem sizes. Dashed lines indicate the ideal curves.
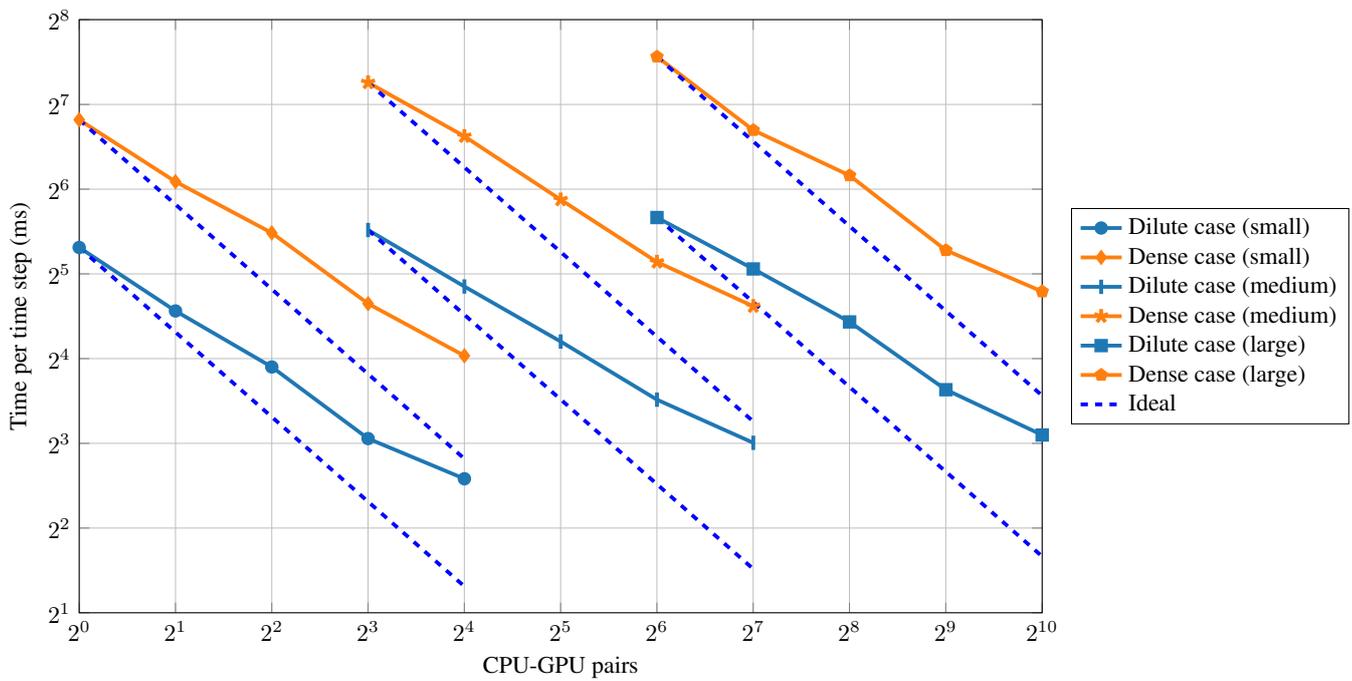
\begin{figure*}
  \centering
  \begin{tikzpicture}[scale=0.9]
    \begin{semilogxaxis}[
        xlabel=CPU-GPU pairs,
        ylabel=Time per time step (ms),
        log basis x={2},
        xmin=1,
        xmax=1024,
        ymode=log,
        log basis y={2},
        ymin=2^1,
        ymax=2^8,
        legend style={at={(1.03,0.5)},anchor=west},
        legend cell align={left},
        width=0.9*\textwidth,
        height=0.4*\textheight,
        grid
        ]
    \addplot[mark=*, mark size=2pt, ultra thick, matplotlibBlue] plot coordinates {
      (1,39.7032)
      (2,23.6182)
      (4,14.9395)
      (8,8.31498)
      (16,5.9862)
    };
    \addplot[mark=diamond*, mark size=2pt, ultra thick, matplotlibOrange] plot coordinates {
      (1,112.8384)
      (2,68.0132)
      (4,44.7182)
      (8,25.0752)
      (16,16.3686)
    };

    \addplot[mark=|, mark size=3pt, ultra thick, matplotlibBlue] plot coordinates {
      (8,45.791)
      (16,28.851)
      (32,18.39274)
      (64,11.43116)
      (128,8.03226)
    };
    \addplot[mark=star, mark size=3pt, ultra thick, matplotlibOrange] plot coordinates {
      (8,153.0142)
      (16,98.5738)
      (32,58.6966)
      (64,35.2456)
      (128,24.5442)
    };

    \addplot[mark=square*, mark size=2pt, ultra thick, matplotlibBlue] plot coordinates {
      (64,50.7164)
      (128,33.309)
      (256,21.5926)
      (512,12.40308)
      (1024,8.56194)
    };
    \addplot[mark=pentagon*, mark size=2pt, ultra thick, matplotlibOrange] plot coordinates {
      (64,188.972)
      (128,103.6524)
      (256,71.5496)
      (512,38.8016)
      (1024,27.6876)
    };

    \addplot[blue, ultra thick, dashed] coordinates { (1,39.7032) (16,39.7032/16) };
    \addplot[blue, ultra thick, dashed] coordinates { (1,112.8384) (16,112.8384/16) };

    \addplot[blue, ultra thick, dashed] coordinates { (8,45.791) (128,45.791/16) };
    \addplot[blue, ultra thick, dashed] coordinates { (8,153.0142) (128,153.0142/16) };

    \addplot[blue, ultra thick, dashed] coordinates { (64,50.7164) (1024,50.7164/16) };
    \addplot[blue, ultra thick, dashed] coordinates { (64,188.972) (1024,188.972/16) };
    \addlegendentry{Dilute case (small)}
    \addlegendentry{Dense case (small)}
    \addlegendentry{Dilute case (medium)}
    \addlegendentry{Dense case (medium)}
    \addlegendentry{Dilute case (large)}
    \addlegendentry{Dense case (large)}
    \addlegendentry{Ideal}
    \end{semilogxaxis}
\end{tikzpicture}
  \caption{\change{Strong scaling performance for both cases and three different problem sizes up to 1024 \gls{cpu}-\gls{gpu} pairs}}\label{fig:strong_scaling}
  \vspace{0.5cm}
\end{figure*}
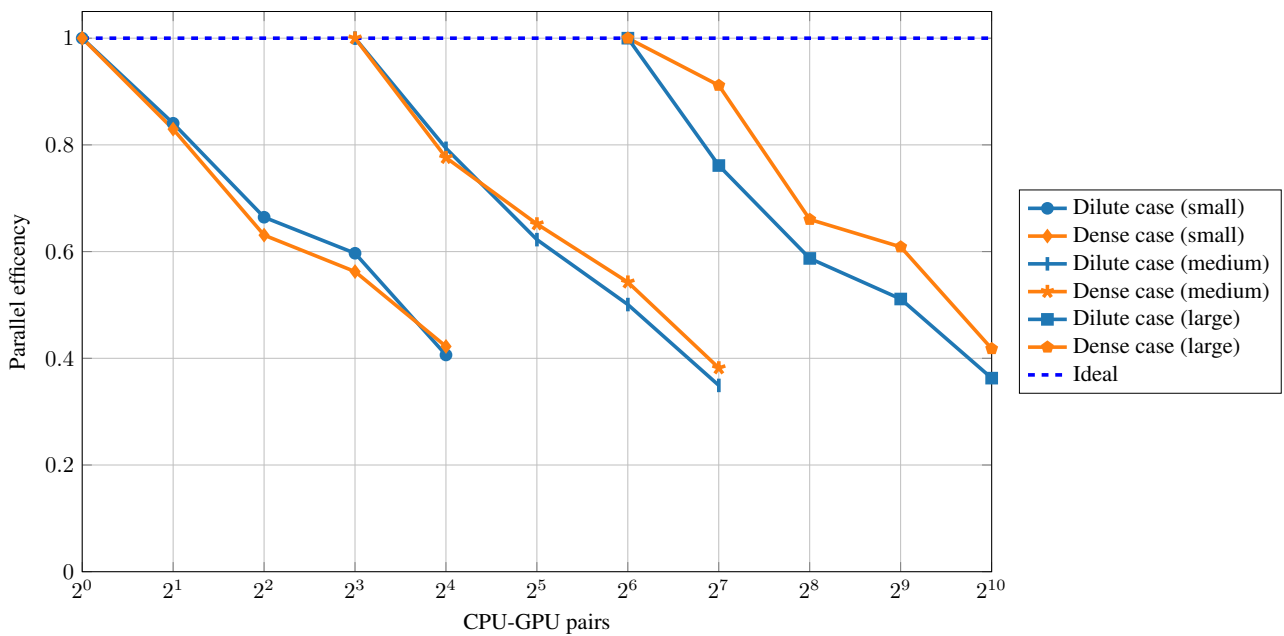
\begin{figure*}
  \centering
  \begin{tikzpicture}[scale=0.85]
    \begin{semilogxaxis}[
        xlabel=CPU-GPU pairs,
        ylabel=Parallel efficency,
        log basis x={2},
        xmin=1,
        xmax=1024,
        ymin=0,
        ymax=1.05,
        legend style={at={(1.03,0.5)},anchor=west},
        legend cell align={left},
        width=0.9*\textwidth,
        height=0.4*\textheight,
        grid
        ]
    \addplot[mark=*, mark size=2pt, ultra thick, matplotlibBlue] plot coordinates {
      (1,1)
      (2,39.7032*0.5/23.6182)
      (4,39.7032*0.25/14.9395)
      (8,39.7032*0.125/8.31498)
      (16,39.7032*0.06125/5.9862)
    };
    \addplot[mark=diamond*, mark size=2pt, ultra thick, matplotlibOrange] plot coordinates {
      (1,1)
      (2,112.8384*0.5/68.0132)
      (4,112.8384*0.25/44.7182)
      (8,112.8384*0.125/25.0752)
      (16,112.8384*0.06125/16.3686)
    };

    \addplot[mark=|, mark size=3pt, ultra thick, matplotlibBlue] plot coordinates {
      (8,1)
      (16,45.791*0.5/28.851)
      (32,45.791*0.25/18.39274)
      (64,45.791*0.125/11.43116)
      (128,45.791*0.06125/8.03226)
    };
    \addplot[mark=star, mark size=3pt, ultra thick, matplotlibOrange] plot coordinates {
      (8,1)
      (16,153.0142*0.5/98.5738)
      (32,153.0142*0.25/58.6966)
      (64,153.0142*0.125/35.2456)
      (128,153.0142*0.06125/24.5442)
    };

    \addplot[mark=square*, mark size=2pt, ultra thick, matplotlibBlue] plot coordinates {
      (64,1)
      (128,50.7164*0.5/33.309)
      (256,50.7164*0.25/21.5926)
      (512,50.7164*0.125/12.40308)
      (1024,50.7164*0.06125/8.56194)
    };
    \addplot[mark=pentagon*, mark size=2pt, ultra thick, matplotlibOrange] plot coordinates {
      (64,1)
      (128,188.972*0.5/103.6524)
      (256,188.972*0.25/71.5496)
      (512,188.972*0.125/38.8016)
      (1024,188.972*0.06125/27.6876)
    };

    \addplot[blue, ultra thick, dashed] coordinates { (1,1) (1024,1) };
    \addlegendentry{Dilute case (small)}
    \addlegendentry{Dense case (small)}
    \addlegendentry{Dilute case (medium)}
    \addlegendentry{Dense case (medium)}
    \addlegendentry{Dilute case (large)}
    \addlegendentry{Dense case (large)}
    \addlegendentry{Ideal}
    \end{semilogxaxis}
\end{tikzpicture}
  \caption{\change{Strong scaling parallel efficiency for both cases and three different problem sizes up to 1024 \gls{cpu}-\gls{gpu} pairs}}\label{fig:strong_scaling_efficiency}
\end{figure*}
The dilute and dense cases exhibit a similar strong scaling behavior for all three problem sizes.
In contrast to the weak scaling, the dense case does not show an overall lower parallel efficiency in the strong scaling.
In all cases, the time per time step decreases when doubling the number of \gls{cpu}-\gls{gpu} pairs. However, the decrease is less than a factor of 2, resulting in a deviation from the ideal curve.
The corresponding parallel efficiency is around 60\% in all cases when doubling the number of \gls{cpu}-\gls{gpu} pairs twice, whereas it drops to about 40\% when doubling the number of \gls{cpu}-\gls{gpu} pairs four times.
The decrease in parallel efficiency is expected in strong scaling since, eventually, the problem size becomes too small to effectively utilize the computational resources while the communication overhead increases.
The most relevant driving force for the decrease in parallel efficiency in the strong scaling originates, similar to the weak scaling analysis (\cref{weak_scaling}), from the frequent synchronizations of the particle properties (\gls{pd} comm).
Similar strong scaling behavior, in particular the decrease in parallel efficiency, has been reported in the literature for other \gls{cfd} codes on different supercomputers~\citep[][]{minExascaleWindEnergy2024, karpLargeScaleDirectNumerical2023}.}

\subsubsection{Potential speedup of hybrid implementations}
We expect that the speedup of the hybrid implementation compared to a \gls{cpu}-only code $S_{\text{hyb}}$ can be estimated based on Amdahl's law as
\begin{equation}
  S_{\text{hyb}}  \approx \frac{1}{1+ \textit{frac}_{\text{acc}} \cdot (\frac{BW_{\text{CPU}}}{BW_{\text{GPU}}}-1)},
  \label{speedup}
\end{equation}
where $BW_{\text{CPU}}$ and $BW_{\text{GPU}}$ are the \gls{cpu} and \gls{gpu} memory bandwidths. $\textit{frac}_{\text{acc}}$ is the \gls{cpu}-only run time fraction of the module accelerated by the hybrid implementation. In our case, this is the \gls{psm} and the coupling. We assume $\textit{frac}_{\text{acc}}$ is memory bound.
We compare the hybrid performance results of this paper with one of the largest \gls{cpu}-only simulations of polydisperse sediment beds~\citep[][]{rettingerRheologyMobileSediment2022}.
The authors conducted the latter simulation on 320 Intel Xeon Platinum 8174 \glspl{cpu}. In total, they computed $2.25\cdot 10^{15}$ lattice cell updates in 48 hours, which leads to a performance of around 41 MLUPs per \gls{cpu} vs. 377 MLUPs per \gls{cpu}-\gls{gpu} pair in the dense case when using 1024 \glspl{gpu}. The latter numbers result in a measured speedup of around 9.2. 
For the Intel Xeon Platinum 8174, we measured a memory bandwidth $BW_{\text{CPU}}$ of 70 GB/s.
The estimated speedup based on \cref{speedup} is around 10.3 (assuming $\textit{frac}_{\text{acc}}=0.95$ and $BW_{\text{GPU}}=1400\ \text{GB/s}$) which is similar to the measured speedup. The latter computation is only a rough estimate since it ignores effects due to different \glspl{cpu}, networks, physical setups, numerical parameters, etc.

\section{Implications and lessons learned}\label{implications}
As a lesson learned, we discovered that the relatively small data transfers between \gls{cpu} and \gls{gpu} can become negligible for coupled applications exhibiting a significant imbalance of the data exchange between the coupled modules and the computations inside the modules in favor of the computations, making hybrid coupling a feasible approach.
This is particularly true for the application used in this work, as only the several orders of magnitude smaller particle data need to be exchanged, not the entire fluid field processed by the \gls{gpu}. As a result, only 0.35\% of the total run time is devoted to \gls{cpu}-\gls{gpu} communication in the dense case.
These findings can also be applied to other applications that exhibit this desired imbalance between computation data size and data transfer between the coupled modules. Other coupled applications, such as a huge flow field around complex deformable geometries like the deformation of blood cells in cellular blood flow, which is referenced in \cref{introduction}, can also exhibit this imbalance. Since the surface data acts as a boundary condition for the deformation simulation, just this surface data needs to be exchanged between the fluid and solid phases.
\change{Approaches with a high ratio of data exchange to computations} (i.e., exchange entire fields between \gls{cpu} and \gls{gpu} in each time step) have been found in the literature to perform undesired on previous architectures, this may change in the future due to new hardware developments.
High bandwidth memory transfer between \gls{cpu} and \gls{gpu} main memory is the trend of architectures like the NVIDIA GH200 Grace Hopper, which makes hybrid implementations promising for more applications (even up to transferring the entire \gls{gpu} data in every time step) and should be further studied in the future.\newline
The advantages of code generation (\cref{fluid_simulation}) are the subject of another lesson learned. We discovered that code generation adds an extra overhead during the implementation stage, which is only beneficial in a framework with long-term support if one depends on highly optimized similar codes, portability to other architectures, and a certain form of expandable, sustainable approach. 
However, because code generation is used, the code functions flawlessly with different \gls{lbm} variations on other modern \gls{gpu} architectures, such as the AMD MI250X, which uses the HIP API, as well as on consumer \glspl{gpu}. Subsequent work will encompass a systematic performance comparison.

\section{Conclusion}\label{conclusion}
On heterogeneous systems, it is pragmatic and, therefore, attractive to use a hybrid parallelization, i.e., different simulation modules running on different hardware.
However, hybrid implementations increase the complexity of achieving good performance and scalability, especially on large-scale systems.
In this paper, we have examined a hybrid coupled fluid-particle simulation with geometrically resolved particles. We use \glspl{gpu} for the fluid dynamics, whereas the particle simulation runs on the \glspl{cpu}.
We have reported and studied the performance of this approach for two cases of a fluidized bed simulation that differ in terms of the number of particles per volume.
The overhead introduced by the hybrid implementation (i.e., \gls{cpu}-\gls{gpu} communication) is negligible because we are transferring only a small amount of data per particle but no fluid cells.
The performance of the fluid simulation is close to utilizing the full memory bandwidth of the A100, implying that using the \gls{gpu} is a good choice for the fluid simulation.
In both cases, the \gls{gpu} routines take most of the run time.
In a weak scaling benchmark, the hybrid fluid-particle implementation reaches a parallel efficiency of 71\% in the dilute case and 53\% in the dense case when using 1024 \gls{cpu}-\gls{gpu} pairs.
The current \gls{pd} methodology requires 32 \gls{cpu}-\gls{cpu} communication steps per time step, which is the driving force for the decrease of the overall parallel efficiency. Our results are limited insofar as different numbers of particle sub-cycles, fluid cells per diameter, etc., will result in different performance results.
We have formulated four criteria that a hybrid implementation must meet to be suitable for the responsible use of heterogeneous supercomputers.
The performance results have shown that our hybrid implementation fulfills all criteria, making it suitable for large-scale simulations on heterogeneous supercomputers.
In the future, we plan to investigate the particle communication steps in more detail regarding the bottleneck and optimization possibilities. We are employing sub-cycles to increase stability for stiff systems. Using other integrators may permit longer time steps and thus less communication due to sub-cycles. We have shown the acceleration potential of hybrid implementations. Therefore, we plan to run coupled fluid-particle simulations of even larger scenarios to better analyze, among others, the physical phenomena of erosion in sediment beds.

\section*{Acknowledgements}
The authors gratefully acknowledge the Gauss Centre for Supercomputing e.V. (\url{www.gauss-centre.eu}) for funding this project by providing computing time on the GCS Supercomputer JUWELS at Jülich Supercomputing Centre (JSC).\newline
The authors gratefully acknowledge the scientific support and HPC resources provided by the Erlangen National High Performance Computing Center (NHR@FAU) of the Friedrich-Alexander-Universität Erlangen-Nürnberg (FAU). The hardware is funded by the German Research Foundation (DFG).

\section*{Declaration of competing interest}
The Authors declare that there is no conflict of interest.

\section*{Funding}
The authors disclosed receipt of the following financial support for the research, authorship, and/or publication of this article: The authors thank the Deutsche Forschungsgemeinschaft (DFG, German Research
Foundation) for funding the project 433735254. The DFG had no direct involvement in this paper; and this work has received funding from the European High Performance Computing Joint Undertaking (JU) and Sweden, Germany, Spain, Greece, and Denmark under grant agreement No 101093393.

\section*{Data availability}
Data is available on Zenodo: \doi{10.5281/zenodo.13951599}.

\section*{Author ORCIDs}
S.\ Kemmler, \url{https://orcid.org/0000-0002-9631-7349};
C.\ Rettinger, \url{https://orcid.org/0000-0002-0605-3731};
U.\ Rüde, \url{https://orcid.org/0000-0001-8796-8599};
P.\ Cuéllar, \url{https://orcid.org/0000-0003-2446-8065};
H.\ Köstler, \url{https://orcid.org/0000-0002-6992-2690};

\section*{Author contributions}
S.\ Kemmler: Conceptualization, Methodology, Software, Validation, Formal analysis, Investigation, Data Curation, Writing - Original Draft, Visualization, Project administration;
C.\ Rettinger: Conceptualization, Writing - Review and Editing;
U.\ Rüde: Writing - Review and Editing;
P.\ Cuéllar: Writing - Review and Editing;
H.\ Köstler: Resources, Writing - Review and Editing, Supervision, Funding acquisition;

\bibliographystyle{SageH}
\bibliography{MyLibrary.bib}

\section*{Author biographies}
\textbf{Samuel Kemmler} is a Ph.D. student at the Chair for System Simulation at the Friedrich-Alexander-Universität Erlangen-Nürnberg (FAU) and a research assistant at the Federal Institute for Materials Research and Testing (BAM) in Berlin. He holds a M.Sc. degree in Computational Engineering and is one of the core developers of the \walberla~HPC framework. His research interests are high-performance computing and particle-resolved sediment transport simulations.\newline

\textbf{Christoph Rettinger} studied Computational Engineering at the FAU. In 2023, he finished his Ph.D. on fully resolved simulation of particulate flows at the Chair for System Simulation.\newline

\textbf{Ulrich Rüde} heads the Chair for System Simulation at the FAU. He studied Mathematics and Computer Science at Technische Universität München (TUM) and the Florida State University. He holds a Ph.D. and Habilitation degrees from TUM. His research interest focuses on numerical simulation and high end computing, in particular computational fluid dynamics, multilevel methods, and software engineering for high performance computing. He is a Fellow of the Society of Industrial and Applied Mathematics.\newline

\textbf{Pablo Cuéllar} studied Civil Engineering at the Universidad Politécnica de Madrid and reiceived his Ph.D. in Civil Engineering from Technical University Berlin in 2011. He is a guest scientist at BAM.\newline

\textbf{Harald Köstler} got his Ph.D. in Computer Science in 2008 on variational models and parallel multigrid methods in medical image processing. 2014 he finished his habilitation on Efficient Numerical Algorithms and Software Engineering for High Performance Computing. Currently, he works at the Chair for System Simulation at the FAU. His research interests include software engineering concepts especially using code generation for simulation software on HPC clusters, multigrid methods, and programming techniques for parallel hardware, especially GPUs. The application areas are computational fluid dynamics, rigid body dynamics, and medical imaging.\newline

\end{document}